\documentstyle[aps, manuscript]{revtex}

\newcommand{\ybco}{YBa$_{2}$Cu$_{3}$O$_{7}$}
\newcommand{\bzo}{BaZrO$_{3}$}
\newcommand{\tc}{$T_{\rm c}$}
\newcommand{\ac}{$A_{\rm c}$}
\newcommand{\aab}{$A_{\rm ab}$}

\begin{document}

\title{Direct Observation and Anisotropy of the Contribution of Gap nodes in the
Low Temperature Specific Heat of \ybco}

\author{Yuxing Wang,  Bernard Revaz, Andreas Erb and Alain Junod.}

\address{Universit\'e de Gen\`eve, D\'epartement de physique de la mati\`ere condens\'ee}
\address{24 quai Ernest-Ansermet, 1211 Gen\`eve 4, Switzerland}

\date{\today}

\maketitle

\begin{abstract}
The specific heat due to line nodes in the superconducting gap of
\ybco$~$ has been blurred up to now by magnetic terms of extrinsic
origin, even for high quality crystals. We report the specific
heat of a new single crystal grown in a non-corrosive \bzo$~$
crucible, for which paramagnetic terms are reduced to
$\approx0.006\%$ spin-1/2 per Cu atom. The contribution of line
nodes shows up directly in the difference $C(B,T)-C(0,T)$ at fixed
temperatures ($T<5~$K) as a function of the magnetic field
parallel to the $c$-axis ($B\le14$ T). These data illustrate the
smooth crossover from $C\propto T^2$ at low fields to $C\propto
TB^{1/2}$ at high fields, and provide new values for gap
parameters which are quantitatively consistent with tunneling
spectroscopy and thermal conductivity in the framework of
$d_{x^2-y^2}$ pairing symmetry. Data for $B$ along the nodal and
antinodal directions in the ab-plane are also provided. The
in-plane anisotropy predicted in the clean limit is not observed.
\end{abstract}


\section{introduction}
Many experiments tend to establish that the symmetry of the order
parameter in high temperature superconductors (HTS) is
$d_{x^2-y^2}$, with a possible minor $s$-wave admixture\cite{1,2}.
Whether they probe the amplitude or the phase of the order
parameter, these measurements have been generally restricted to
the surface of the samples. Alternatively, specific heat ($C$)
experiments have been used to search for bulk evidence of a
non-conventional gap, taking advantage of their sensitivity to the
low energy excitations of a system. In a $d_{x^2-y^2}$
superconductor, the gap $\Delta (\vec{k})$ vanishes and changes
sign on lines of nodes in $k$-space. The finite slope of the gap
at the nodes causes a linear increase of the density of states
(DOS) at low energy, $N(E)\propto|E|$. As the electronic part of
$C/T$ is proportional to the DOS averaged over an interval
$\approx k_BT$ about the Fermi level, it follows that
$C_{electron}/T=\alpha T$ at $T$ $\ll$ \tc$~$ in zero magnetic
field. This contrasts with the exponential law
$C_{electron}/T\propto T^{-2.5}exp(-\Delta_0/T)$ that would apply
at very low temperature to a fully-gapped $s$-wave superconductor.

In a magnetic field, the energy of carriers circulating around a
vortex is shifted by Doppler effect. This allows them to be
excited above the local gap on the Fermi surface near the nodes.
Detailed calculations show that a contribution $C_{vortex}/T =$
\ac $B^{1/2}$ arises in the $T=0$ limit \cite{3}. In a
conventional isotropic $s$-wave superconductor, no significant
contribution would be expected at low temperature from such a
mechanism, since a field on the order of $B_{\rm c2}$ is needed to
shift the energy by an amount comparable to the gap $\Delta_0$.

Two other mechanisms may contribute to the low temperature
specific heat of superconductors in a magnetic field: the Zeeman
shift\cite{4,5} and localized levels in vortex cores.\cite{6} The
Zeeman and Doppler contributions to $C/T$ scale as $B/B_{\rm c2}$
and $(B/B_{\rm c2})^{1/2}$, respectively, so that the latter
dominates at low fields. Localized levels in vortex cores also add
a contribution $\propto (B/B_{\rm c2})$. For a low-\tc$~$
superconductor with a large coherence length $\xi$, the mini-gaps
$\approx \Delta_0/k_F\xi $ between these levels are so small that
the local DOS can be considered as continuous; the average DOS is
proportional to the area occupied by vortex cores and entails $C/T
\propto (B/B_{\rm c2})$. This is the main field-induced
contribution at low temperature for classic superconductors. For
HTS, the localization in smaller vortex cores increases the
separation between levels. Tunneling spectroscopy\cite{7} tends to
show that the mini-gaps are then on the order of
$\approx0.3k_B$\tc, which implies that the contribution of core
levels should be negligible in the temperature range $T<~$\tc /20
investigated here. Furthermore, the very existence of localized
core levels for a pure $d$-wave superconductor is questionable, as
they can leak along the nodal directions and mix with other
states.

Summarizing, the presence of line nodes in HTS is reflected by
terms $C(T \ll$ \tc,$~B=0)/T=\alpha T$ and $C(T=0, B\ll B_{\rm
c2})/T=~$\ac $B^{1/2}$, whereas for a fully gapped, low
temperature superconductor $C(T\ll$ \tc,$~B=0)/T\propto
\exp(-\Delta_0/T)\approx 0$ and $C(T=0, B\ll B_{\rm c2})/T\propto
B$. However these simple criteria are not easily applied in
practice because of the presence of additional contributions to
the measured specific heat. In YBa$_{2}$Cu$_{3}$O$_{7-\delta}$,
the main contributions arise from lattice vibrations, weakly
interacting paramagnetic centers, and a linear term $C_{\rm
lin}=\gamma(0)T$ of uncertain origin. All three are
sample-dependent. In order to separate the $d$-wave contribution
of interest, the Stanford\cite{8,9} and Berkeley\cite{10,11}
groups fitted models to their data, and concluded that a $d$-wave
contribution was present. Numerical results were: $\alpha= 0.10$
to $0.11$ mJ/K$^{3}$mol, \ac$~$= $0.88$ to $0.91$
mJ/K$^{2}$T$^{1/2}$ mol according to Moler et al.,\cite{8,9} who
used both twinned and untwinned crystals, and $\alpha= 0.064\pm
0.02$ mJ/K$^3$mol, \ac$~= 0.91$ {mJ/K}$^2$T$^{1/2}$mol according
to Wright et al.,\cite{11} who used ceramic samples. In this
paper, we report low temperature specific heat measurements for a
twinned and fully oxidized \ybco$~$ crystal in higher fields. The
concentration of magnetic impurities for this new crystal is so
low that it is possible to see for the first time in
YBa$_{2}$Cu$_{3}$O$_{7-\delta}$ the contribution due to line nodes
in the raw data prior to any correction or fit. Most uncertainties
due to the background are avoided by looking at the difference
between the specific heat in a magnetic field and that in zero
field, $C_{\rm dif}(B,T)=C(B,T)-C(0,T)$, which is electronic in
origin. We obtain $\alpha= 0.21$ mJ/K$^3$mol$\pm 20\%$ and
\ac$~=1.34$ mJ/K$^2$T$^{1/2}$mol$\pm3\%$. Together with recent
results from thermal conductivity,\cite{12}
photoemission,\cite{13} and tunneling spectroscopy,\cite{14} these
new values allow a numerically consistent picture of the
anisotropy of the order parameter for \ybco$~$ to be constructed.
A second issue that can be studied for the first time owing to the
low magnetic contamination is the anisotropy of the vortex
specific heat, which attracted recently theoretical
interest.\cite{15}

\section{experimental results}
1. Sample

The sample used for the present measurement is a 18 mg
YBa$_{2}$Cu$_{3}$O$_{7.00}$ single crystal (code AE429G)
flux-grown in a \bzo$~$ crucible. The latter non-reactive material
allows a slow growth with little contamination.\cite{16} Final
chemical purity is better than 99.995\%.\cite{17} Large crystals
have to be screened because of possible CuO/BaCuO$_{2}$ flux
inclusions. The effect of this second phase is potentially huge
since only 0.9\% BaCuO$_{2}$ in weight would double the heat
capacity of the sample at 3 K, with an intricate field
dependence.\cite{18} The present crystal, about $2\times1.5\times1
~{\rm mm}^3$ in size, is twinned, with its smallest dimension
along the $c$-axis. Detwinning was not attempted, as it was found
that standard procedures used for crystals of that size lead to a
degradation of their properties in terms of transition width and
concentration of magnetic centers. The sample was fully oxygenated
in 100 bar O$_{2}$, 330C, for 200 hours. The calorimetric
transition midpoint, \tc$~$= 87.8 K, is typical of overdoped
YBa$_{2}$Cu$_{3}$O$_{7.00}$. The unextrapolated peak-to-peak
amplitude of the $C/T$ anomaly at \tc$~$ is $61$ \rm mJ/K$^2$mol,
i.e. 4.5\% of the essentially phononic background (Fig. \ref{1}).
This amplitude is reproducible for crystals grown in \bzo$~$(cf.
18 mg crystal AE195G\cite{19} and 40 mg crystal AE276G\cite{20}).
Fully oxidized samples are preferred for the present purpose for
two reasons: clusters of oxygen vacancies effectively act as
residual impurities,\cite{21} introducing scattering that could
modify drastically the shape of the DOS at $E=0$ in a $d$-wave
superconductor,\cite{22} and a correlation has been observed
between the concentration of oxygen vacancies and the amplitude of
the Schottky anomaly,\cite{23} which tends to mask the $d$-wave
contribution.

2. Specific heat: method and tests.

The low temperature specific heat was measured for $1.2\le T< 5\rm
K$ and $0\le B \le 14\rm T$ using a thermal relaxation
technique.\cite{23} For calibration purpose, we first measured in
$B = 0$ two silver samples of 6N purity with masses 9 and 21 mg.
This allowed us to determine independently the heat capacity of
the addenda (sapphire sample holder, contributing part of the
phosphor bronze suspension wires, epoxy, carbon film
heater/thermometer, silicone grease for sample mounting) and the
specific heat of silver, which could be then compared with
reference data. The heat capacity of the smaller Ag sample was on
the same order of magnitude as that of the \ybco$~$ crystal in the
middle of the temperature range. We verified that the measured
specific heat of Ag after subtraction of the addenda did not
change significantly with the field. Silver, which has a small
nuclear magnetic moment, presents over copper standards the
advantage of a negligible nuclear heat capacity in the temperature
and field range investigated here. Figure \ref{2} shows the
measured low temperature specific heat of silver, together with
reference data.\cite{24} The insert of Fig. \ref{2} shows the
Sommerfeld constant $\gamma$ and Debye temperature $\Theta_{\rm
D}$ obtained in fields $B$ = 0, 4, 8 and 14 T. The results are
consistent with accepted values $\Theta_{\rm D}=226~\rm K$ and
$\gamma = 0.65~\rm mJ/K^2mol$.\cite{24} Note that for \ybco, the
Debye specific heat $C\propto \Theta_{\rm D}^{-3}$ is 6.7 times
smaller (per gram-atom), so that the determination of $\gamma(B)$
is more accurate.

Specific heat data can be measured both with increasing and with
decreasing temperature. The former are obtained from the response
to switching on the heater, the latter to switching off. The heat
capacity is given in both cases by:

\begin{equation}
[R(T(t))-R(T(\infty ))]I^2=C(T(t))\frac{dT}{dt} +\int_{T(\infty
)}^{T(t)}k(T^{'})dT^{'}
\label{eq1}
\end{equation}
where $R(T)$ is the resistance of the carbon film used both as a
heater and a thermometer, $I$ is the current through the film, set
to a constant value at time $t>0$, $C(T)$ is the total heat
capacity at temperature $T$, $T(\infty)$ is the final asymptotic
temperature for a given value of $I$ (measured typically after ten
time constants), $k(T)$ is the thermal conductivity of the
supporting wires, measured separately when $dT/dt=0$ using various
heater powers and various base temperatures. We use the
approximations $T(t)=(T_{\rm n}+T_{\rm n+1})/2$ and $dT/dt =
(T_{\rm n+1}-T_{\rm n})/(t_{\rm n+1}-t_{\rm n})$ with 50 ms time
intervals between the $n^{\rm th}$ and $(n+1)^{\rm th}$ data
acquisition, therefore a single relaxation provides numerous
independent $C(T)$ points. These data tend to accumulate at the
end of a relaxation until the error on $(dT/dt)^{-1}$ diverges; we
discard data with $|T_{\rm n+1}-T_{\rm n}|<$ 4 mK. Such
accumulations are seen in Fig. \ref{2} and Fig. \ref{3} near $T^2
= 2,5,11$ and $22~\rm K^2$. The up/down procedure allows one to
check that measurements are independent of the heating rate (i.e.,
there is no "$\tau_2$ effect" in the sense of Ref. \cite{25}). On
a fine scale, there is a time dependence of the specific heat of
\ybco$~$ at the highest fields and lowest temperatures. This is
due to the hyperfine specific heat of Cu nuclei, which are weakly
coupled to the electrons\cite{27}. Data at $T<1.5$ K in $B=14$ T
are discarded for this reason.

3. Specific heat of the \ybco$~$crystal.

Figure \ref{3} shows the low temperature specific heat of sample
AE429G in different magnetic fields, shown separately for
$B\|$[001] along the $c$-axis (Fig. \ref{3}a), for $B\|$[110]
along the $ab$-diagonal where a gap node is expected (Fig.
\ref{3}b), and for $B$ along both the $a$- and $b$-axes (the
sample is twinned), i.e. along the antinodal direction (Fig.
\ref{3}c). For the sake of simplicity, the latter direction is
called $B\|$[100]. These are raw data, i.e. only the heat capacity
of the sample holder is subtracted, otherwise no correction is
performed for possible BaCuO$_{2}$/CuO impurities originating from
flux,\cite{18} paramagnetic centers with $S=1/2$ and/or
$S=2$,\cite{26} nuclear hyperfine contribution,\cite{27} etc. In a
first approximation the specific heat is essentially the sum of
two dominant contributions. The $C_{\rm ph}\cong \beta T^3$
lattice term shows up as a field-independent slope in the $C/T$
versus $T^2$ plot, whereas the field-dependent electronic term
$C_{\rm el}\cong \gamma (B)T$ shifts the curves parallel to each
other.

The Schottky effect of residual paramagnetic centers causes a
low-temperature upturn in low fields, and a characteristic maximum
of $C_{\rm Sch}/T$ at $T_{\rm max}[{\rm K}]=0.415B[\rm T]$ for
$S=1/2$ spins. At high fields, the amplitude of the maximum
decreases whereas its width increases, so that the Schottky
contribution finally merges into the background. By fitting the
expression

\begin{equation}
C_{\rm sch}=\frac{Nz^{2}e^{z}}{(e^{z}+1)^2},z=\frac{2gS\mu _{\rm
B}B_{eff}}{k_{\rm B}T}, B_{eff}=\sqrt{B_{ext}^2+B_{int}^2}
\label{eq2}
\end{equation}
we find $N = 0.12\pm 0.02~\rm mJ/K.gat$, i.e. 0.005 to 0.007\%
spins-1/2 per Cu atom, and $gS=1.1$ to 1.15.\cite{28} The
amplitude $N$ and the value of $gS$ are determined for $B=4$ or 8
T along the $ab$-plane, in order to minimize the relative vortex
contribution and to locate the maximum of the Schottky peak in a
convenient temperature range. In this fit, which only serves to
characterize the Schottky contribution, we have assumed that the
other terms are $\beta T^3$ for phonons and $\gamma(B)T$ for
electrons (a more general analysis is given in Section 3).
Imposing then the same value of $N$ for $B=0$ or 0.5 T, we obtain
an equivalent interaction field $B_{\rm int} \approx$ 0.2 to 0.5
T. The Schottky term is small relative to the vortex term for
$B\|c$, but is significant for $B\bot c$ at intermediate fields.
In particular, it explains why the raw data at 14 T do not lie
clearly above those in 8 T in Fig. \ref{3}b and Fig. \ref{3}c, in
spite of the monotonically increasing vortex term. In any case,
high fields improve the ratio of the vortex specific heat over the
Schottky specific heat.

Extrinsic magnetic contributions for sample AE429G are unusually
small. This is not an artifact. Measurements performed in our
laboratory on other samples, including ceramics\cite{23} and
detwinned single crystals from various sources have shown a more
usual typical concentration of $\approx 0.1\%$ spin-1/2 centers
per copper atom. Such "impurities" make it difficult to separate
all contributions, in particular at low fields where magnetic
interactions between spins cannot be neglected.

The present measurements are compared with earlier work on
YBa$_{2}$Cu$_{3}$O$_{7- \delta}~$ single crystal\cite{8} in
Fig.\ref{b2}. The same fields and the same presentation as for
Fig.3 of Ref\cite{8} are used, i.e. a $\beta T^3~$ term is
subtracted. he present data differ in the following respects: the
measurement range is shifted by a factor of roughly two toward low
temperature where the phonon contribution is smaller; the $\beta$
coefficient is $\sim 20\%$ smaller (possibly owing to hardenng of
Cu-O chain at full oxygenation, but also to a smaller high-order
$electronic$ contribution); the magnetic Schottky term causing
deviations from a horizontal line in non-zero field is suppressed
by a factor of ~15; the residual linear term $\gamma(0)$ is $\sim
30\%$ smaller (but similar to that of sample U1 in Ref.\cite{9}).
These features, which can be summarized by saying that all
non-$d$-wave terms contribute less, facilitate the analysisand
give more weight to the $d$-wave specific heat for the present
crystal.

\section{discussion}

1. Field along the $c$-axis.

In order to get a premilinary insight into the final results, we
plot again the data of Fig.\ref{3}(a) ($B\|c$)$~$in
Fig.\ref{3}(d), after having subtracted the 8 Tesla curve taken as
a reference. The remaining specific heat qualitatively shows the
evolution of the field-dependent electronic contribution $C_{\rm
e}/T$. Anticipating the discussion, we expect the high field
curves to be free of the anomalous $C\propto T^2$ term, hence the
choice of the 8 Tesla curve as a reference. Higher fields would
only introduce more scatter. This plot only involves smoothing of
the reference curve (residuals are shown by the 8 T data set), but
neither parameter fit nor correction of any kind. Fig.\ref{3}(d)
evidences the parallel shift of the curves in high fields, $C_{\rm
e}/T = f(B)$, and the progressive appearance of a positive slope
at very low fields, ending into a $C_{\rm e}/T\propto T$ term in
$B = 0$. The small upturn at low temperature in the low field
curves, together with the faint maximum barely observable at
intermediate fields, can be both explained by the presence of a
small residual Schottky contribution. At the highest fields and
lowest temperatures, data obtained from separate relaxations at
different velocities dT/dt do not join smoothly because of the
weak coupling of electrons with Cu nuclei.\cite{27}

The analysis of a difference $C(T, B)-C(T, B_{\rm ref})$
circumvents the problem of modeling the field-independent
background (lattice specific heat, residual "linear term", etc,).
However, in the present status of knowledge a field -independent
sub-linear or logarithmic term is no more justified tha a linear
one, so that we avoid any fit of the background. The following
discussion will only depend on differences. However, for a
comparison with theory, $B_{\rm ref}=0$ is more convenient as a
reference than $B_{\rm ref}=$ 8 T. Therefore we shall concentrate
on $C_{\rm dif}(T,B)=C(T,B)-C(T,0)$.

Figure \ref{4} shows $C_{\rm dif}(B)$ at fixed temperatures
$T=2,~3$, and $4$ K, prior to correction for the magnetic Schottky
effect. It is immediately apparent that the vortex specific heat
increases approximately with the square root of the field. This
behavior can be readily explained by existing theory, up to the
deviations from the square root law. We briefly recall the
essential results.

Volovik et al.\cite{3,29,30} first pointed out that in the mixed
state of a superconductor with line nodes, supercurrents around a
vortex core cause a Doppler shift of the quasiparticle excitation
spectrum. If the superfluid velocity is $\vec{v}_s$, the
quasiparticle excitation spectrum $E(\vec{k})$ is shifted by
$\vec{k}\cdot \vec{v}_s$ . This shift has important effects around
nodes, where its value is comparable to the width of the
superconducting gap. The density of states at the Fermi level is
strongly affected:

\begin{equation}
N(0)=n\int \frac{d^3k}{(2\pi )^3} \int d^2r \delta
(E(\vec{k},\vec{r})-\vec{k} \cdot \vec{v}_s(\vec{r}))
\label{eq3}
\end{equation}
The average superfluid velocity depends on the inverse intervortex
distance, $\langle v_s \rangle \propto 1/R(B)$, so that the
integral is proportional to $R(B)\propto 1/B^{1/2}$. For $B\gg
B_{\rm c1}$, the number of vortices $n$ is proportional to $B$, so
that $N(0)\propto B^{1/2}$ and $C_{\rm el}/T\propto B^{1/2}$ at
$T\to 0$. Thus one has two regimes, depending on whether the
thermal energy $k_{\rm B}T$ is large or small compared to the
typical Doppler energy, one that is quadratic in $T$ ($C_{\rm
el}=\alpha T^2$) in zero field at $T\ll~$\tc, and one that is
linear in $T$ ($C_{\rm el}=$\ac$ TB^{1/2}$) at zero temperature
and $B\ll B_{\rm c2}$. More precisely, for a weak-coupling
superconductor with $d_{x^2-y^2}$ symmetry, K\"ubert et
al.\cite{31} and Vekhter et al.\cite{15,32} obtained:

\begin{equation}
\frac{C_{el,[001]}}{\gamma_{\rm
n}T}=(\frac{8}{\pi})^{1/2}(\frac{B}{B_{\rm c2}/a^2})^{1/2}+
\frac{14(2\pi )^{1/2}}{15}(\frac{k_{\rm B}T}{\Delta
_0})^2(\frac{B_{\rm c2}/a^2}{B})^{1/2}+..., \frac{TB_{\rm
c2}^{1/2}}{T_{\rm c} B^{1/2}} \ll 1
\label{eq4}
\end{equation}

\begin{equation}
\frac{C_{el,[001]}}{\gamma_{\rm n}T}=\frac{27\zeta (3)}{\pi
^2}\frac{k_{\rm B}T}{\Delta _0}+ \frac{3ln2}{2\pi }\frac{\Delta
_0}{k_{\rm B}T}\frac{B}{B_{\rm c2}/a^2}+..., \frac{TB_{\rm
c2}^{1/2}}{T_{\rm c} B^{1/2}} \gg 1
\label{eq5}
\end{equation}
where $\zeta(3)=1.202$... and $a$ is a constant of order unity
depending only on the vortex lattice geometry. The latter constant
is defined differently in Ref. \cite{31} and \cite{15,32}; we have
chosen here the notation of Refs. \cite{12} and \cite{31}$~$(see
note \cite{33}). These equations may be rewritten in terms of the
Fermi velocity $v_{\rm F}$ to avoid any reference to the
ill-defined upper critical field of HTS. In this form, they are no
longer restricted to the weak-coupling case. Keeping only the
leading terms, we have then:

\begin{equation}
\frac{C_{el,[001]}}{\gamma_{\rm n}T}=\frac{4a}{\Phi
_0^{1/2}}\frac{\hbar v_{\rm F}}{\pi \Delta_0}B^{1/2}+...,
\frac{\Phi_0^{1/2}k_{\rm B}}{\hbar v_{\rm F}} \frac{T}{B^{1/2}}
\ll 1
\label{eq6}
\end{equation}

\begin{equation}
\frac{C_{el,[001]}}{\gamma_{\rm n}T}=\frac{27\zeta (3)}{\pi
^2}\frac{k_{\rm B}}{\Delta_0}T+..., \frac{\Phi_0^{1/2}k_{\rm
B}}{\hbar v_{\rm F}} \frac{T}{B^{1/2}}\gg 1 \label{eq7}
\end{equation}
These asymptotic formulas extrapolate to the same result at a
crossover temperature $T_{\rm cross}(B)$ given by:

\begin{equation}
T_{cross}(B)=\frac{(2\pi)^{3/2}} {27\zeta (3)}
 \frac{\Delta_0}{k_{\rm
B}}(\frac{B}{B_{\rm c2}/a^2})^{1/2}
\label{eq8}
\end{equation}
The ratio between the experimental parameters \ac$~$and $\alpha$
depends only on the Fermi velocity:

\begin{equation}
\frac{A_{\rm c}}{\alpha}
 \equiv
 \frac{lim_{T\to 0}(C/TB^{1/2})}
      {lim_{B\to 0}(C/T^2)}=
 \frac{T_{cross}(B)} {B^{1/2}}=
 \frac{4\pi \hbar}{27\zeta (3)\Phi_0^{1/2}k_{\rm B}}
 av_{\rm F}
 \label{eq9}
\end{equation}
The full function across the crossover regime has not been
calculated analytically, but has the following scaling
property:\cite{30,34}

\begin{equation}
\frac{C_{el,[001]}}{\gamma_{\rm n}T}(\frac{B_{\rm
c2}}{B})^{1/2}=F_{c,[001]}(x)
\label{eq10}
\end{equation}
where $x\equiv T/T_{\rm cross}(B)$.\cite{33} The empirical
interpolating function

\begin{equation}
F_{c,[001]}(x)\cong (\frac{8}{\pi})^{1/2}a(1+x^2)^{1/2}
\label
{eq11}
\end{equation}
allows one to map the full specific heat function (Fig. \ref{5}).
Far above the crossover temperature, the Doppler shift can be
neglected, so that only the bulk term $C/T\propto T$ arising from
the V-shape of the DOS survives. Conversely, far below $T_{\rm
cross}(B)$, only the field-dependent plateau in the DOS at $E\to
0$ is probed by thermal excitations, so that the Doppler term
$C/T\propto B^{1/2}$ dominates. In any case one has assumed
$T\ll~$\tc$~$and $B\ll B_{\rm c2}$, which is satisfied in the
present data limited to $T<5$ K and $B\le 14$ T. As we shall see
below, $T_{\rm cross}/B^{1/2}\cong 6.4~{\rm K/T}^{1/2}$ when
$B\|c$, so that measurements of $C_{\rm dif}(T, B)$ from 2 to 4 K
in fields from 0.16 to 14 T probe the region $0.085\le x\le 1.6$,
which spans both the high field/low temperature limit and the
crossover regime $x\approx 1$. This implies that both constants
\ac$~$ and $\alpha$ can be determined independently from the data.

Since we are interested in the difference $C_{\rm dif}$, the
relevant scaling function is $F_{\rm dif}(x)$ defined by:

\begin{equation}
\frac{C_{dif,[001]}}{\gamma_{\rm n}T}=(\frac{B}{B_{\rm
c2}})^{1/2}F_{c,[001]}(x)-\frac{27\zeta (3)k_{\rm B}}{\pi ^2
\Delta_0 }T\equiv (\frac{B}{B_{\rm c2}})^{1/2}F_{dif,[001]}(x)
\label{eq12}
\end{equation}
and the corresponding interpolation function becomes:

\begin{equation}
F_{dif,[001]}(x)\cong (\frac{8}{\pi})^{1/2}a[(1+x^2)^{1/2}-x]
\label{eq13}
\end{equation}
Figure \ref{6} shows the scaling plot $C_{\rm dif}/TB^{1/2}$
versus $T/B^{1/2}$ corresponding to Eq.(\ref{eq12}). The data for
$B\|c$, measured at fixed temperatures $T = 2, 3$ and 4 K, and now
corrected for the Schottky contribution(in the main frame, not in
the insert), collapse onto a single curve, thus supporting the
existence of line nodes. This test also shows that the additional
energy scale introduced by possible impurity scattering is
negligible compared to the other scales determined by thermal
smearing and Doppler shift. Strong impurity scattering would lead
to a breakdown of the scaling property, with $N(0)$ decreasing
ultimately as $|B\log B|$ rather than $B^{1/2}$ at low
fields.\cite{31} A comparison with numerical calculations (Fig. 4
of Ref. \cite{31}) shows that our data lie essentially in the
clean limit. This also implies that the observed residual linear
term $\gamma(0)T$, with $\gamma(0)\approx 15\%$ of the
normal-state Sommerfeld constant $\gamma_{\rm n}$, is mostly not
caused by impurity scattering.

Figure \ref{7} shows how parameters \ac$~$ and $\alpha$ can be
extracted directly from the data. We plot $C_{\rm dif}/T$ versus
$B^{1/2}$ at $T$ = 2, 3 and 4K. At high fields, data fall on
parallel lines. This is the region $x\ll 1$ where

\begin{equation}
\frac{C_{dif,[001]}}{T}\mid _{x\ll 1}
 \cong 1.596a
\frac{\gamma_{\rm n}B^{1/2}} {B_{\rm c2}^{1/2}}
 -3.288
 \frac{\gamma_{\rm n}k_{\rm B}} {\Delta_0}
 T \equiv {\rm A}_c B^{1/2}-\alpha T
 \label{eq14}
\end{equation}

The slope determines the leading field-dependent term with \ac$=
1.34~\rm mJ/K^2T^{1/2}mol\pm 3\%$, whereas the parallel shift with
changing temperature in the high field limit determines the bulk
term with $\alpha =0.21~\rm mJ/K^3mol\pm 20\%$. Their ratio yields
$av_{\rm F}\cong 1.0\times 10^7$ cm/s. Together with the maximum
gap width $\Delta_0= 20~$meV determined by scanning tunneling
spectroscopy,\cite{14} we further obtain $\gamma_{\rm n}\cong
15~\rm mJ/K^2mol$ and $B_{\rm c2}/a^2\cong 310$ T. These results
agree with the re-evaluation by K\"ubert and Hirschfeld of earlier
data characterized by various scattering rates.\cite{31} The
crossover temperature is given by $T_{\rm cross}/B^{1/2}=~$\ac$
/\alpha =6.4~{\rm K}/T^{1/2}$. The parameters \ac$~$ and $\alpha$
are substantially larger than those determined earlier and
recalled in the Introduction. This may be due in part to the use
of a crystal that is overdoped rather than optimally doped.
Scattering by impurities is probably lower.\cite{31} In addition,
the methods used by different authors have different sensitivities
to various sources of experimental errors. The present results
allow one to construct a scenario in the framework of
$d_{x^2-y^2}$ pairing symmetry that is consistent with other types
of experiments such as tunneling spectroscopy and thermal
conductivity. Following Chiao et al.\cite{12} we assume a
cylindrical Fermi surface; the \ac$~$ parameter can be expressed
in terms of the slope $v_2\equiv
\partial \Delta/\partial p_{\bot}$ of the gap on the Fermi
surface in the direction perpendicular to the line of nodes:

\begin{equation}
A_{\rm c}=\frac{8k_{\rm B}^2}{3\hbar \Phi_0^{1/2}}
\frac{V_{mol}}{d} \frac{a}{v_2}
\label{eq15}
\end{equation}
This follows from Eq.~\ref{eq6}, with $\gamma _{\rm n}=(\pi
^2/3)k_{\rm B}^2N(0)V_{\rm mol}$, $N(0)=m^{\ast}/(\pi \hbar ^2d)$
where $d=c/2=0.584$ nm is the average interlayer distance, $V_{\rm
mol}=104.6$ cm$^3$/mol the molar volume, $\Delta (\phi )=\Delta _0
\cos (2\phi )$, and $|d\Delta (\phi )/d\phi |_{\phi =\pi /4} =
\hbar k_{\rm F}v_2=2\Delta _0$ at the node. Eq. (15) yields
$v_2/a\cong 1.4\times 10^6$ cm/s. In the same approximation,
$\alpha$ is given by

\begin{equation}
\alpha =\frac{18\zeta (3) k_{\rm B}^3}{\pi \hbar ^2}
\frac{V_{mol}}{d} \frac{1}{v_2 v_{\rm F}}
\label{eq16}
\end{equation}
which yields $v_2v_{\rm F}\cong 1.4\times 10^{13}~\rm (cm/s)^2$.
Up to this point, the geometrical parameter $a$ is undetermined.
The recent analysis of thermal conductivity in the universal low
temperature regime by Chiao et al.\cite{12} determined $v_{\rm
F}/v_2=14\pm 20\%$ in YBa$_{2}$Cu$_{3}$O$_{\approx 7}$. Combining
this result with ours one obtains $v_{\rm F}\cong 1.4\times 10^7$
cm/s, $v_2\cong 1.0\times 10^6$ cm/s, $a\cong 0.70$, $B_{\rm
c2}\cong 150$ T, corresponding to a coherence length $\xi \cong
15~{\AA}$. The geometrical parameter is of order one as
anticipated, and the upper critical field is consistent with
previous estimations.\cite{35,36} The Fermi velocity is smaller
than determinations based on angle-resolved photoemission
spectroscopy (ARPES), $v_{\rm F}\cong 2.5\times 10^7$
cm/s.\cite{13} The difference appears to be real, even taking into
account the large uncertainty margin on this parameter (Table I),
and may be due to a contribution from chain states with a low
Fermi velocity in our overdoped sample. In this respect it is
interesting to note that YBa$_{2}$Cu$_{4}$O$_{8}$, which contains
completely filled double chains, presents a large negative
curvature in the plot of $C/T$ versus $T^2$ in $B=0$ at $T<5$
K.\cite{23} If this curvature is attributed to a bulk $d$-wave
$\alpha T^2$ term, then $\alpha$ is still three times larger (per
chain) for YBa$_{2}$Cu$_{4}$O$_{8}$ than for \ybco, and
consequently the product $v_2v_{\rm F}$ is three times smaller,
showing a potential source of variations of $v_{\rm F}$. The Fermi
wave number is $k_{\rm F}=2\Delta_0/\hbar v_2\cong0.6~
{\AA}^{-1}\cong 3/4~\pi /a_1$ ($a_1\cong 3.85~{\AA}$ is the
lattice constant). The latter result is consistent with the
quasi-2D hole band centered on the S-point ($\vec{k} =[\pi
/a_1,\pi /b,0]$) found by band structure calculations\cite{37} and
by ARPES,\cite{13} which show that plane states cross the Fermi
surface near $k_{\rm F}\cong 0.69$ to $0.75~{\AA}^{-1}$ along the
nodal direction and $k_{\rm F}\cong 0.58$ to $0.64~{\AA}^{-1}$
along the antinodal directions. The overall consistency provides
strong support to the quasi-2D, $d_{x^2-y^2}$-wave model. In
particular using a normalized gap function $\Delta(\phi)/\Delta_0$
similar to that of Bi$_{2}$Sr$_{2}$CaCu$_{2}$O$_{8}$ does not lead
to any contradiction.

Based on the previous data, one can further evaluate the Fermi
temperature $T_{\rm F}=\hbar k_{\rm F}v_{\rm F}/2k_{\rm
B}=(\Delta_0/k_{\rm B})(v_{\rm F}/v_2)\cong 3300$ K and the mass
enhancement factor $m^{\ast}/m_{\rm e}=\hbar k_{\rm F}/m_{\rm
e}v_{\rm F}\cong 5$. Note that the latter ratio depends only on
the product of experimental quantities $\Delta_0\alpha $. It is
interesting to note that the Bose-Einstein condensation
temperature of such heavy quasi-2D hole pairs, if preformed above
\tc, should occur at\cite{38}

\begin{equation}
T_{\rm BE}= \frac{2\pi\hbar^2 nd}{k_{\rm B}m^{\ast}}
\frac{1}{\ln(\frac{2k_{\rm B}T_{\rm BE} \Gamma^2
m^{\ast}d^2}{\hbar^2})} \cong 100~{\rm K}
\label{eq17}
\end{equation}
where $\Gamma = 5.3$ is the anisotropy ratio of the upper critical
field.\cite{39,40} For the pair density, we used the value
$n=1.0\times 10^{21} cm^3$ obtained by scaling the amplitude of
the $\lambda$ anomaly of the specific heat of $^4$He at 2.18 K,
for which the boson density is $n_b=2.2\times 10^{22}~$cm$^3$, to
the $\lambda$-anomaly of \ybco$~$ at \tc. The fact that $T_{\rm
BE}$ is close to \tc$~$ may be a coincidence for \ybco, since the
penetration depth calculated from these parameters exceeds
experimental values by a factor of about two. Additionally, the
ratio\cite{12} $\Delta _0/E_{\rm F}=2/(\pi k_{\rm F}\xi)\cong (
v_2/v_{\rm F})\cong 1/14$ shows that only a small part of the
carriers near the Fermi surface are paired.

Finally the initial assumption that vortex core levels do not
contribute significantly to the specific heat at low temperature
can be tested $a~posteriori$. Self-consistent calculations of the
electronic structure of a vortex line in NbSe$_{2}$ have shown
that the lowest energy excitation depends on temperature, with a
value at $T\ll~$\tc$~$of the order of $E_{1/2}\cong
{\Delta_0}^2/E_{\rm F}$.\cite{41} Applying the same criterion to
\ybco$~$ leads to $E_{1/2}/k_{\rm B}\cong 17$ K, about half the
estimation based on STS,\cite{7} but large enough to be neglected
in the specific heat below 4 K.

The above determinations are summarized in Table \ref{tabl1}. In
order to evaluate uncertainty margins, we included the error on
experimental parameters \ac, $\alpha$, and $v_{\rm F}/v_2$, and
neglected that on $\Delta_0$.

2. $B\| ab$-plane

The main motivation of the measurements for $B\bot c$ was the
prediction of a variation of $\gamma (T=0, B, \phi )$ with the
azimutal angle $\phi$ between the field and the $a$-axis.\cite{15}
The qualitative physics underlying this effect may be understood
by noticing that when a field is applied along the antinodal [100]
direction, all four nodes contribute equally to the DOS, with an
amplitude proportional to $4\cos(\pi /4)$, whereas for $B$ along
the nodal direction [110], the Doppler effect vanishes for those
particles which travel parallel to the field, so that only two
nodes contribute with an amplitude proportional to $2\cos(0)$.
Quantitative estimations depend on details of the theory, so that
a 30\% variation appears as an upper limit for a clean, tetragonal
crystal with a 2D band. Such fourfold oscillations of $\gamma
(T=0, B, \phi )$ versus $\phi$, if observed, would provide a
robust test of $d$-wave symmetry. Moler et al.\cite{9} did not
observe any such oscillation within experimental resolution. A
second motivation is the experimental verification of the general
features of the vortex specific heat predicted theoretically for
$B\bot c$\cite{15,32}.

Data analysis for $B\bot c$ is more involved than for $B\| c$
(compare Fig. \ref{8} and Fig. \ref{4}). Even for the present
crystal, the residual Schottky contribution due to paramagnetic
centers is no longer negligible compared to the reduced vortex
contribution. We face for the $B\bot c$ configuration the same
problem as the Stanford\cite{8,9} and Berkeley\cite{10,11} groups
did for the $B\|c$ configuration. As data for $B\bot c$ are
expected to lie in the crossover regime (the crossover temperature
scales with $(B/\Gamma)^{1/2}$), a fit to a $C=$\aab$ TB^{1/2}$
law is not justified beforehand. We resort then to a scaling plot
which is more generally valid, first using raw data (Fig.
\ref{9}a; note that the scaling variable is $B^{1/2}/T$ in this
plot, so that unlike Fig. \ref{6} high fields are on the right).
At this stage, data do not collapse on a common curve. In the same
figure we plot the Schottky specific heat of spin-1/2 paramagnetic
centers, with a concentration equivalent to 0.006\% of the Cu
atoms. The position of the peaks corresponds to the maxima of the
raw data, confirming that the paramagnetic centers have a spin
$S=1/2$. Note that the Schottky correction is negligible in $B=14$
T. In the scaling plot of Fig. \ref{9}b, we show the data after
having subtracted the Schottky contribution. The parameters of the
Schottky anomaly were adjusted manually to obtain the best
collapse onto a common curve. In this operation, the most
efficient parameter is the concentration of paramagnetic centers;
only minor improvements are obtained by refining the gyromagnetic
ratio and the interaction field. Data define a wide high field
region ($T/B^{1/2} < 2~{\rm K}/T^{1/2}$) where
$C=~$\aab$TB^{1/2}$, \aab$=0.18~\rm mJ/K^2T^{1/2}$mol $\pm 10\%$,
therefore \ac$/$\aab$\cong 7.4$. The latter determination does not
depend critically on the subtraction of the Schottky anomaly; it
suffices to consider only 14 T data, $B\bot c$, where the Schottky
contribution can be neglected, and compare with 14 T data, $B\| c$
(Fig. \ref{8} and Fig. \ref{4}).

The ratio \ac/\aab$~$is larger than the square root of the
anisotropy ratio at \tc, $\Gamma^{1/2} = 2.30\pm 0.07$. This is
expected, at least qualitatively. We recall the results of Ref.
\cite{15,32} for the specific heat with the field parallel to the
planes, assuming moderate anisotropy, cylindrical Fermi surface
and $\Delta_{\rm k}=\Delta_0 \cos (2\phi )$:

\begin{equation}
\frac{C_{el}}{\gamma_{\rm n}T}=
(\frac{2}{\pi})^{1/2}(\frac{B/\Gamma}{B_{\rm c2}/a^2})^{1/2} +
 \frac{27\zeta (3)}{2\pi^2}
 \frac{k_{\rm B}T}{\Delta_0} +..., B\|[110],
 \frac{T (\Gamma B_{\rm c2})^{1/2}}{T_{\rm c} B^{1/2}} \ll 1
\label{eq18}
\end{equation}

\begin{equation}
\frac{C_{el}}{\gamma_{\rm n}T}=
\frac{2}{\pi^{1/2}}(\frac{B/\Gamma}{B_{\rm c2}/a^2})^{1/2} +
\frac{28\pi^{1/2}}{15}(\frac{k_{\rm B}T}{\Delta_0})^2
(\frac{B_{\rm c2}/a^2}{B/\Gamma})^{1/2}+..., B\|[100],
\frac{T (\Gamma B_{\rm c2})^{1/2}}{T_{\rm c}B^{1/2}} \ll 1
\label{eq19}
\end{equation}

\begin{equation}
\frac{C_{el}}{\gamma_{\rm n}T}=
\frac{27\zeta (3)}{\pi^2}
\frac{k_{\rm B}T}{\Delta_0} +
\frac{3\ln 2}{4\pi}
\frac{\Delta_0}{k_{\rm B}T}
\frac{B/\Gamma}{B_{\rm c2}/a^2}+...,
B\|[110],B\|[100], \frac{T(\Gamma B_{\rm c2})^{1/2}}{T_{\rm c}
B^{1/2}} \gg 1 \label{eq20}
\end{equation}
As before, we have restored the convention for the geometrical
parameter $a$ of Ref.~\cite{31}; see Note\cite{33}. In the high
temperature regime, one recovers the bulk isotropic $C\propto T^2$
behavior, irrespective of the field direction. At $T=0$, the
expected ratio between the $c$-axis and nodal specific heat is
$C(0, B\| [001])/C(0, B\|[110])=2\Gamma ^{1/2}\cong 4.6$, that
between the $c$-axis and antinodal specific heat is $C(0, B\|
[001])/C(0, B\|[100])=(2/\Gamma)^{1/2}\cong 3.3$. The theoretical
parameter-free curves given by the second-order approximations,
Eq.~\ref{eq18}-\ref{eq20}, are plotted in Fig. \ref{9}b. The order
of magnitude is correct, but Eq.~\ref{eq18}-\ref{eq20} first fail
to reproduce the wide plateau for $B^{1/2}/T>0.5~\rm T^{1/2}/K$,
second tend to overestimate the vortex specific heat for large
values of $B^{1/2}/T$, and third predict an in-plane anisotropy
for large values of $B^{1/2}/T$ that is not observed. There is no
way to improve the agreement by changing the spin and the
amplitude of the Schottky correction, which is negligible for
large values of $B/T$.

One is tempted to conclude that the anisotropy ratio at low
temperature is larger than the value $\Gamma=5.3$ found at
\tc$~$(an example of such behavior is given by 2H-NbSe$_{2}$ for
which $(B_{\rm c2,ab}/B_{\rm c2,c})=3.4$ at $T\to 0$ whereas
$(dB_{\rm c2,ab}/dT)/(dB_{\rm c2,c}/dT)=2.7$ at \tc),\cite{42} but
this does not remedy the situation. A larger anisotropy would
indeed decrease the coefficient of the vortex term \aab, and
therefore would shift down the theoretical curves; however this
would leave the coefficient of the bulk term $\alpha$ unchanged
and, remembering that the zero field specific heat is subtracted
in $C_{\rm dif}$, this bulk term would cause a positive slope for
all values of $B^{1/2}/T$ plotted in Fig.~\ref{9}b. Therefore, the
plateau would not be reproduced.

The experimental points show that the difference between the
in-field and the zero-field specific heat contains essentially a
small vortex term \aab$TB^{1/2}$ for $T/B^{1/2}<2~\rm K/T^{1/2}$,
apparently without any $-\alpha T^2$ contribution from the bulk
term. Leaving aside the disturbing scenario $\alpha=0$, one
concludes that the application of a field $B\bot c$ leaves the
$\alpha T^2$ term unchanged at low temperature, so that it cancels
in $C_{\rm dif}$. In terms of density of states, a small V-shape
dip subsists at $E_{\rm F}$ in the middle of a field-induced
plateau that moves up with $B^{1/2}$. Within this picture, the
sharp crossover seen experimentally near $T/B^{1/2}=2~\rm
K/T^{1/2}$ corresponds to the edge of the dip, not to \aab$/\alpha
\cong 0.86~\rm K/T^{1/2}$. Such a fine structure of the DOS could
indicate that the anisotropic London theory leading to
Eq.~\ref{eq18}-\ref{eq20}, and which is valid for moderate
anisotropy, does not take into account all aspects of vortex
physics for $B\bot c$.

The absence of in-plane anisotropy is a second puzzling feature.
It is already apparent in the raw data, Fig.~\ref{3}b and
Fig.~\ref{3}c, and more precisely seen in the scaling plot of
Fig.~\ref{9}b, where the points for $B$ along the antinodal
direction hardly differ from those for $B$ along the nodal
direction. In order to perform a more quantitative analysis, we
focus on the $B=14$ T data for two reasons: first the Schottky
correction is negligible, and second the vortex contribution
follows a simple high field $d$-wave law, as just discussed. By
fitting the parameters of a model $C/T=\gamma (B)+\beta T^2$ to
the data in $B=14$ T (Table \ref{tabl2}), we find
$\gamma(B\|[100])-\gamma (B\|[110]) = 0.04\pm 0.06~\rm mJ/K^2mol$.
The residuals show no structure suggesting any missing term. The
sign is correct but the amplitude is an order of magnitude smaller
than the $d$-wave prediction, \ac $B^{1/2}[(2\Gamma
)^{-1/2}-(4\Gamma)^{-1/2}]=0.45~\rm mJ/K^2mol$ at 14 T in the
clean limit. The 8 T data confirm the 14 T result. In order to
make this comparison as significant as possible, we took care not
to add or remove any addendum mass (in particular the adhesive)
when rotating the sample at room temperature. The experimental
reproducibility is documented by the data in $B=0$, which were
measured separately in each one of the three sample positions
[001], [100] and [110] along the magnet axis (Table \ref{tabl2}).

In order to reconcile the lack of any significant variation of
$\gamma(\phi )$ with the existence of $d$-wave pairing, various
arguments may be invoked.\cite{32} First, the 2D model of the DOS
may be too crude. Numerical estimations have shown that the
amplitude of the oscillations of $\gamma (T=0, B, \phi )$ with
$\phi$ is reduced in the 3D case.\cite{32} Second, the full 30\%
effect develops only at T = 0; distinct crossover effects decrease
the in-plane anisotropy at any finite temperature,\cite{15} but
this reduction is taken into account in the model curves of Fig.
\ref{9}b. Third, orthorhombicity shifts the sharp minimum of
$\gamma (T=0, B, \phi =\pi /4)$ slightly off the nodal
direction.\cite{32} Together with twinning, this replaces the
single minimum by a double dip centered on $\phi=\pi /4$ with
reduced depth. Scattering may smooth the oscillations;\cite{43} we
believe that this is not the main cause here since scattering
would also have led to a breakdown of the $\gamma \propto B^{1/2}$
dependence for $B\|c$. Finally, the implications of the in-plane
anisotropy of the penetration depth in \ybco, $\lambda_{\rm
a}>\lambda_{\rm b},$\cite{44} were not addressed. In short,
\ybco$~$does not appear to be suitable for a quantitative test of
$d$-wave pairing symmetry based on $ab$-plane anisotropy, owing to
its orthorhombicity.

\section{conclusion}

The low contamination of the present crystal by paramagnetic
impurities has allowed to observe directly a field-induced
contribution to the specific heat of \ybco$~$that can be accounted
for by the theoretical treatment of a $d$-wave superconductor with
a quasi-2D band at the Fermi level. The parameters inferred from
this experiment have a typical $\pm 20\%$ accuracy as they depend
critically on the value of the bulk $T^2$-term of the zero-field
specific heat. Together with the results of tunneling
spectroscopy, photoemission and thermal conductivity, these
results allow an overdetermined and consistent set of microscopic
parameters to be established.

The bulk confirmation of $d$-wave properties provided by specific
heat is important, as stressed by previous studies on this
subject. However experiment and theory still indicate some
unsolved puzzles. First recall that a fully gapped s-wave
component, if present, would not give rise to any measurable
contribution at $T\ll \Delta_0/k_{\rm B}$, so that the present
experiments cannot provide any information on this topic. It has
been pointed out that non-linear variations of $\gamma (B)$ with
the field also occur in s-wave superconductors, both from an
experimental\cite{42,45,46} and theoretical\cite{47,48} point of
view. The present experiments probably cannot distinguish between
$C\propto B^{1/2}$ and $C\propto B^{0.41}$ as proposed in Ref.
\cite{48}. The expected $ab$-plane anisotropy could not be
observed. As the present results leave little hope of obtaining a
decisive answer using \ybco$~$as a working material, tetragonal
compounds should rather be investigated, preferably with a Ba-free
composition in order to avoid the presence of BaCuO$_{2}$
impurities. The field dependence of the specific heat for $B\bot
c$ remains to be explained in detail. Finally, it has been pointed
out that the relatively large local DOS that should be present
between the vortex cores in the Doppler model has not been
observed up to now by tunneling spectroscopy in \ybco,
notwithstanding several attempts\cite{49}. Therefore, in spite of
considerable progress, a full understanding of the mixed state of
\ybco$~$has yet to be reached. Bulk investigations of
unconventional pairing in other HTS, if they can be prepared with
the same degree of purity, would be most informative.

\section{appendix}

In a previous article, referred to as BR,\cite{50} we have
estimated by a different method the vortex specific heat of
another crystal (AE37G), also grown in \bzo, with a larger
Schottky anomaly corresponding to 0.03\% spin-1/2 per Cu atom. The
field-independent terms were canceled by considering the
anisotropic component $C(B\| c)-C(B\bot c)$ rather than the
difference $C(B)-C(0)$. It was assumed that anisotropy enters only
through the ratio $B_{\rm c2,ab}/B_{\rm c2,c}=5.3$, an estimation
that now has to be revised. BR obtained \ac$~=1.8~\rm
mJ/K^2T^{1/2}mol$, and, noticing that $[C(B\|c)-C(B\bot
c)]/TB^{1/2}$ was not constant, concluded that data were taken in
the crossover regime.

Now that the specific heat for $B\bot c$ has been measured
separately, we can check these conclusions. The present finding is
that the specific heat for $B\bot c$ is much smaller than
expected. Rather than $C(B\|c)/C(B\bot c)\propto (B_{\rm
c2,ab}/B_{\rm c2,c})^{1/2}\cong 2.3$ as used in BR,
$C(B\|c)/C(B\bot c)\cong 7.4$ over a wide field range. Therefore
$C(B\|c)-C(B\bot c)$ does not contain $1-\Gamma ^{-1/2}=57\%$ of
$C_{\rm vortex}(B\|c)$, but $1-1/7.4=87\%$. Our previous result,
based on a measured anisotropic component \ac$-$\aab$=1.0~\rm
mJ/K^2T^{1/2}mol$ at $T\to 0$, has to be rescaled to \ac$=1.2~\rm
mJ/K^2T^{1/2}mol$. The difference with the present result,
\ac$=1.34~\rm mJ/K^2T^{1/2}mol$, is essentially due to the way
data were extrapolated to $T=0$ in Fig.~3 of BR.

The second point is the crossover temperature that separates the
low T, high B regime from the high T, low B regime. In a plot of
$[C(B\| c)-C(B\bot c)]/TB^{1/2}$ versus $T/B^{1/2}$, the crossover
is the value of $T/B^{1/2}$ at which $[C(B\| c)-C(B\bot
c)]/TB^{1/2}$ extrapolates linearly to zero, starting from its
initial value at $T/B^{1/2}=0$ (see e.g. Fig. \ref{6}). In view of
the smallness of the vortex contribution for $B\bot c$, one has
approximately $C(B\| c)-C(B\bot c)\approx C(B\| c)-C(0)$. Fig. 2
of BR, redrawn on a linear scale, shows that the crossover
estimated in this way is \ac$ /\alpha =5$ to $7~{\rm K}/T^{1/2}$
for $B\|c$, consistent with the result for the present crystal
AE429G, \ac$ /\alpha =6.4~{\rm K}/T^{1/2}$. In the high field
region $0<T/B^{1/2}\ll$ \ac$ /\alpha$, thus below the crossover
temperature, $C(B\|c)/TB^{1/2}$ is constant but $[C(B\|
c)-C(0)]/TB^{1/2}$ must decrease linearly with $T/B^{1/2}$, a
point that was not made clear in BR.

\vspace*{0.5cm}

\centerline{\bf {Acknowledgements}}
\vspace*{0.5cm}

This work was supported by the Fonds National Suisse de la
Recherche Scientifique. The authors thank I. Vekhter, L.
Taillefer, G. E. Volovik, E. Schachinger, A. Holmes and A. Manuel
for fruitful discussions.

\vspace*{2cm}
After completion of this work, we became aware of
recent developments in theory, both in the Doppler (Volovik)
approach\cite{51} and in self-consistent calculations of the
electronic structure of a $d_{x^2-y^2}$ vortex\cite{52}. Although
the shape of the envelope is preserved, the density of states at
the Fermi level in the latter approach is always zero, on a very
fine energy scale, irrespective of the field\cite{53,54,55}. This
is not in contradiction with the present results in the 1-4 K
range, which are broadened by thermal smearing.

\begin{figure}[]
\caption{Total specific heat of the present crystal (AE429G) and
another crystal also grown in \bzo$~$ (AE195G) near \tc. The
critical temperature is defined here as the point with the largest
negative slope of $C/T$ versus $T$.} \label{1}
\end{figure}

\begin{figure}[]
\caption{Specific heat $C/T$ versus $T^2$ of the silver test
sample. Full line: standard reference data.\protect\cite{24}
Insert: fitted Debye temperature and Sommerfeld constant in
different magnetic fields.} \label{2}
\end{figure}

\begin{figure}[]
\caption{Comparison of the present data(series (b) on the left
side 1.2 - 2.4K). As for Fig.\ref{3} of Ref.\protect\cite{8}, the
specific heat is plotted per mol units as $(C- \beta T^3)/T$ vs
$T$, $\beta=0.392 {\rm mJ/molK}^2$ for the data of Ref.
\protect\cite{8}, and $\beta=0.305 {\rm mJ/molK}^2$ for the
present work. The magnetic field is applied perpendicular to the
planes. The same subset of fields is shown.} \label{b2}
\end{figure}

\begin{figure}[]
\caption{Low temperature specific heat $C/T$ of \ybco$~$(AE429G)
for different fields, raw data. (a), $B\|[001]$, from bottom to
top  $B$ = 0, 0.16, 0.5, 1, 2, 3, 4, 6, 8, 10, 12, 14 T. (b),
$B\|[110]$, $B$ = 0, 0.5, 1, 2, 4, 8, 14 T. (c), $B\|[100]$ and
$[010]$, $B$ = 0, 8, 14 T. (d), same data as (a), after
subtraction of the $B$ = 8T curve}
\label{3}
\end{figure}

\begin{figure}[]
\caption{Specific heat $C/T$ versus the field $B~(B\|c)$ at fixed
temperatures, raw data, showing a nearly square-root law.}
\label{4}
\end{figure}

\begin{figure}[]
\caption{Model calculation of the low temperature specific heat
versus temperature and field $(B\|c)$ in the presence of line of
nodes, showing the high-field, low temperature regions where
$C/T\propto B^{1/2}$ and the high temperature, low field regions
where $C/T\propto T$. The parameters correspond to those for the
present crystal AE429G.}
\label{5}
\end{figure}

\begin{figure}[]
\caption{d-wave scaling plot of the specific heat difference
$C_{\rm dif}/TB^{1/2}$ versus $T/B^{1/2}$ $(B\|c)$. A Schottky
contribution has been subtracted. Full line: interpolated scaling
function. Insert: the same plot without correction for the
Schottky contribution.}
\label{6}
\end{figure}

\begin{figure}[]
\caption{Plot of the specific heat difference $C_{\rm dif}/T$
versus $B^{1/2}$ $(B\|c)$ at fixed temperatures, allowing one to
extract the parameters \ac$~$and $\alpha$. A Schottky contribution
has been subtracted. Full lines: interpolated scaling function.
Insert: the same plot without correction for the Schottky
contribution.} \label{7}
\end{figure}

\begin{figure}[]
\caption{Plot of the specific heat difference $C_{\rm dif}/T$
versus $B~(B\|[110])$ at fixed temperatures, raw data.}
\label{8}
\end{figure}

\begin{figure}[]
\caption{(a), same raw data as for Fig. 9 $(B\|[110])$ in a
scaling plot $C_{\rm dif}/TB^{1/2}$ versus $B^{1/2}/T$. The full
lines show the estimated Schottky contribution at different
temperatures. (b), remaining vortex contribution after having
removed the Schottky anomaly, both for $B\|[110]$ and $B\|[100]$
Dotted and dashed lines: anisotropic $d$-wave model (see text).}
\label{9}
\end{figure}

\begin{table}
\begin{tabular}{lcl}
  sample                        & AE429G                 &  \\
  $V_{\rm mol}=13V_{\rm gat}$   & 104.6                  &cm$^{3}$/mol \\

  $M_{\rm mol}=13M_{\rm gat}$   & 666.2                  & g/mol \\

  \tc                           &87.8 $\pm$ 0.05         &K\\

  $\Gamma$ \cite{40}            &5.3 $\pm$ 0.3           &   \\

  \ac                           &1.34 $\pm$ 3 $\%$
  &mJ/K$^2$T$^{1/2}$mol\\

  \aab                          &0.18 $\pm$ 10 $\%$
  &mJ/K$^2$T$^{1/2}$mol\\

  $\alpha$                      &0.21 $\pm$ 20 $\%$       &mJ/K$^3$mol\\

  $v_2/a$                       &1.42 $\times$ 10$^6\pm$ 3 $\%$
  &cm/s\\

  $v_{\rm F}v_2$               &1.4 $\times$ 10$^{13}\pm$ 20 $\%$
  &(cm/s)$^2$\\

  $av_{\rm F}$                  &1.0 $\times$ 10$^7\pm$ 24 $\%$
  &cm/s\\

  $a^2v_{\rm F}/v_2$            &6.9 $\pm$ 27 $\%$        &   \\

  $T_{\rm crossover}/B^{1/2}$   &6.4 $\pm$ 24 $\%$
  &K/T$^{1/2}$\\

  $\Delta_0$\cite{7}            &20                       &meV\\

  $\gamma_n$                    &15 $\pm$ 20 $\%$

  &mJ/K$^2$mol\\

  $B_{\rm c2}/a^2$              &310 $\pm$ 11 $\%$         &T\\

  $m^{\ast}/m_e$                &5.1 $\pm$ 20 $\%$         &   \\

  $v_{\rm F}/v_2$\cite{12}      &14 $\pm$ 20 $\%$          &   \\

  $a$                           &0.70 $\pm$ 23 $\%$        &   \\

  $v_{\rm F}$                   &1.4 $\times$ 10$^7\pm$ 53 $\%$
  &cm/s\\

  $v_2$                         &1.0 $\times$ 10$^6\pm$ 27 $\%$
  &cm/s\\

  $B_{\rm c2}$                  &150 $\pm$ 68 $\%$         &T\\

  $\xi$                         &15 $\pm$ 30 $\%$          &{\AA}\\

  $k_{\rm F}$                   &0.61 $\pm$ 27 $\%$        &{\AA}$^{-1}$\\

  $T_{\rm F}$                   &3300 $\pm$ 20 $\%$        &K\\
\end{tabular}
\centering \caption{\ybco$~$parameters obtained from the
field-induced specific heat. Rows below $\Delta_0$ make use of the
gap value from tunneling spectroscopy.\protect\cite{7} Rows below
$v_{\rm F}/v_2$ make use of this ratio from thermal
conductivity.\protect\cite{12} See text for definitions. The
effect of experimental uncertainties on \ac, $\alpha$ and $v_{\rm
F}/v_2$ is indicated; the error on $\Delta_0$ is neglected.}
\label{tabl1}
\end{table}

\pagebreak

\begin{table}
\begin{tabular}{cccccc}
  $B$(T)  &orientation      &$\gamma(B)~[\rm mJ/K^2mol]$   & $\Theta_{\rm D}~$(K)
  &r.m.s. error [mJ/K$^2$mol]     &number of data\\

  0   &[001]           &2.19  &422.3   &0.035   &357 \\

  0   &[100]+[010]     &2.17  &423.0   &0.042   &647 \\

  0   &[110]           &2.20  &423.4   &0.025   &305 \\

  8   &[100]+[010]     &2.79  &427.4   &0.047   &514 \\

  8   &[110]           &2.75  &426.7   &0.039   &301 \\

  14  &[100]+[010]     &2.96  &425.6   &0.049   &210 \\

  14  &[110]           &2.92  &425.6   &0.057   &723 \\
\end{tabular}
\vspace*{0.5cm}
  \centering
  \caption{Results of fits of the high field specific heat for $B\bot c$:
  coefficient of the linear term $\gamma (B)$ and Debye temperature $\Theta_{\rm D}$. The
  Schottky correction is fixed and corresponds here to 0.007\%
  spin-1/2 per Cu atom with $g=2.2$ and $B_{\rm int}=0.5~$T. The data in $B=0$ with 3
  orientations with respect to the magnet axis are shown only to document
  the reproducibility, since the bulk $C\propto T^2$ $d$-wave term is not included
  in the fit (attempts to include it yield values that are strongly
  correlated with $B_{\rm int}$ and $\Theta_{\rm D}$ and do not improve significantly
  the fit).}
  \label{tabl2}
\end{table}


\begin{references}
\bibitem{1} D. Scalapino, Phys. Rev. 250, 329 (1995).
\bibitem{2} D. Pines and P. Monthoux, J. Phys. Chem. Solids 56, 1651 (1995).
\bibitem{3} G.E. Volovik, JETP Lett. 58, 469 (1993).
\bibitem{4} B. M\"uhlschlegel, Z. Physik 155, 131 (1959).
\bibitem{5} Kun Yang and S.L. Sondhi, Phys. Rev. B 57, 8566 (1998).
\bibitem{6} C. Caroli, P.G. de Gennes and J. Matricon, Phys. Lett. 9, 307 (1964).
\bibitem{7} I. Maggio-Aprile, Phys. Rev. Lett. 75, 2754 (1995).
\bibitem{8} K.A. Moler, D.J. Baar, J.S. Urbach, R. Liang, W.N. Hardy, and A. Kapitulnik,
 Phys. Rev. Lett. 73, 2744 (1994).
\bibitem{9} K.A. Moler David L. Sisson, Jeffrey S. Urbach, Malcolm R. Beasley, Aharon Kapitulnik
 et al., Phys. Rev. B 55, 3954 (1997).
\bibitem{10} R.A. Fisher et al., Physica C 252, 237 (1995).
\bibitem{11} D.A. Wright, J.P. Emerson, B.F. Woodfield, J.E. Gordon,
 R.A. Fisher and N.E. Phillips, Phys. Rev. Lett. 82, 1550 (1998).
\bibitem{12} May Chiao, R.W. Hill, Christian Lupien, Bojana Popic, Robert Gagnon and
 Louis Taillefer Phys. Rev. Lett. 82, 2943 (1999).
\bibitem{13} M.C. Schabel, C.-H. Park, A. Matsurra, Z.-X. Shen, D.A. Bonn,
 R. Liang and W.N. Hardy, Phys. Rev. B 57, 6090 and 6107 (1998).
\bibitem{14} I. Maggio-Aprile, Ch. Renner, A. Erb, E. Walker, and O. Fischer,
 Phys. Rev. Lett. 62, 214 (1995).
\bibitem{15}I. Vekhter, P.J. Hirschfeld, J.P. Carbotte and E.J. Nicol, in
 "Physical Phenomena at High Magnetic Fields - III",
 edited by Z. Fisk, L. Gor'kov and R. Schrieffer, World Scientific,
 Singapore 1999, p. 410.
\bibitem{16} A. Erb, J.-Y. Genoud, F. Marti, M. D\"aumling, E. Walker and R. Fl\"ukiger
 J. Low Temp. Phys. 105, 1023 (1996).
\bibitem{17} A. Erb, E. Walker and R. Fl\"ukiger, Physica C 245, 245 (1995).
\bibitem{18} J.-Y. Genoud, A. Mirmelstein, G. Triscone, A. Junod, and J. Muller,
 Phys. Rev. B 52, 12833 (1995).
\bibitem{19} A. Junod, M. Roulin, J.-Y. Genoud, B. Revaz, E. Wlaker, A. Erb, C. Marcenat,
 R. Calemczuk and F. Bouqet, Physica C 275, 245 (1997).
\bibitem{20} A. Junod, A. Erb and C. Renner, Physica C 317, 333 (1999).
\bibitem{21} A. Erb, A.A. Manuel, M. Dhalle, F. Marti, J.-Y. Genoud, B. Revaz, A. Junod, D.
 Vasumathi, S. Ishibashi, A. Shukla, E. Walker, Ø. Fischer, R. Flükiger, R. Pozzi, M.
 Mali and D. Brinkmann, Solid State Commun. 112, 245 (1999).
\bibitem{22} K. Maki and H. Won, Ann. Phys. (Leipzig) 5, 320 (1996).
\bibitem{23} A. Junod, in: "Studies of High Temperature Superconductors,
 vol. 19", ed. A. Narlikar, Nova Science Publishers, Commack, New York 1996, p. 1.
\bibitem{24} G.T. Furukawa, W.G. Saba and M.L. Reilly, Nat. Bur. Stand. Ref. Data Ser.
 (USA), No. 18, p.1 (1968).
\bibitem{25} R. Bachmann, F.J. Disalvo, T.H. Geballe, R.L. Greene, R.E. Howard, C.N. King,
 H.C. Kirsch, K.N. Lee, R.E. Schwall, H.-U. Thomas and R.B. Zubeck  Rev. Sci. Instrum. 43,
 205 (1972).
\bibitem{26} J.P. Emerson, D.A. Wright, B.F. Woodfield, J.E. Gordon,
 R.A. Fisher and N.E. Phillips, Phys. Rev. Lett. 82, 1546 (1999).
\bibitem{27} The informed experimenter might be puzzled by the missing upturn at the
lowest temperatures ($<2\rm K$) and highest fields ($>10\rm T$).
The absence of the Cu nuclear specific heat is due to the short
characteristic time of our measurement (0.6-0.8s at 1.2 K and 14
T, $B\|c$), to be compared with Korringa's electron-nucleus
thermalization time, $\tau_1T\sim$1.1 sK. This causes (only at
high field and low temperature) a time dependence of the heat
capacity; data at the end of a relaxation when $|$d$T/$d$t|\to0$
lie systematically above those obtained at the beginning of a
relaxation when $|$d$T/$d$t|$ is large (e.g. 0.2-0.4 K/s near 1.3
K). Generally, for a given temperature, we have data taken at both
slow and fast rates. By discarding the former we ensure that the
nuclear contribution of Cu is in large part eliminated. The same
reasoning holds for alternating current specific heat methods;
above a few Hertz, Cu nuclei no longer follow the oscillation of
the electronic temperature.
\bibitem{28} F. Mehran, S.E. Barnes, E.A. Giess and T. R. McGuire,
 Solid State Commun. 67, 55 (1988).
\bibitem{29} N.B. Kopnin and G.E. Volovik, JETP Lett. 64, 641 (1996).
\bibitem{30} G.E. Volovik, JETP Lett. 65, 491 (1997).
\bibitem{31} C. K\"ubert and P.J. Hirschfeld, Solid State Commun. 105, 459 (1998).
\bibitem{32} I. Vekhter, P.J. Hirschfeld, J.P. Carbotte and E.J. Nicol,
 Phys. Rev. B 59, R9023 (1999).
\bibitem{33} In the definition of the geometrical parameter $a$, we have used the
convention of K\"ubert and Hirschfeld \cite{31} to facilitate
comparison with Ref. \cite{12}. The convention used by Vekhter et
al.\cite{15,21} is different and ensures that for $a=1$ there is
one flux quantum per unit cell of the vortex lattice approximated
by a circle of radius $R=(\Phi_{0}/\pi B)^{1/2}$. The equivalence
is $a\equiv a_{K\ddot{u}bert}=(\pi^{1/2}/2)a_{Vekhter}$. The
Doppler energy scales are $E_{H,K\ddot{u}bert}=a\Delta_{0}
(B/B_{\rm c2})^{1/2}$ in K\"ubert's convention and
$E_{H,Vekhter}=(a_{Vekhter}\hbar v_{F}/2)(\pi B/\Phi_{0})^{1/2}$
in Vekhter's convention, so that $E_{H,K\ddot{u}bert}=(2/\pi )^{
1/2}E_{H,Vekhter}$. The crossover temperature at which the high
field leading term of the specific heat $C = ATB^{1/2}$ becomes
equal to the bulk term $C=\alpha T^2$ corresponds to
$k_BT/E_{H,K\ddot{u}bert}= 0.485$, to $k_BT/E_{H,Vekhter}=0.387$,
and to $x=[27\zeta (3)/(2\pi )^{3/2}]k_BT/E_{H,K\ddot{u}bert}=
[27\zeta(3)/(4\pi )]k_BT/E_{H,Vekhter}=1$ in the present work. Of
course, physical results do not depend on these conventions.
\bibitem{34} S.A. Simon and P.A. Lee, Phys. Rev. Lett. 78, 1548
(1997); Phys. Rev. Lett. 78, 5029 (1997).
\bibitem{35} U. Welp, W.K. Kwok, G.W. Crabtree, K.G. Vandervoort,
 and J.Z. Liu  Phys. Rev. Lett. 62, 1908 (1989).
\bibitem{36} G. Triscone, A.F. Khoder, C. Opagiste, J.-Y. Genoud, T. Graf,
 E. Janod, T. Tsukamato, M. Couach, A.Junod, J. Muller,  Physica C 224, 263 (1994).
\bibitem{37} O.K. Andersen, O. Jepsen, A.I. Liechtenstein, and I.I. Mazin,
 Phys. Rev. B 49, 4145 (1994).
\bibitem{38} R. Micnas, J. Ranninger and S. Robaszkiewicz, Rev. Mod. Phys.
62, 113 (1990).
\bibitem{39} M. Roulin, A. Junod and E. Walker, Physica C 260, 257 (1996).
\bibitem{40} M. Roulin, B. Revaz, A. Junod, A. Erb and E. Walker in: "Physics
and Materials Science of Vortex States,
Flux Pinning and Dynamics", eds. R. Kossowsky, S. Bose, V. Pan and
Z. Durusoy, NATO Science Series E: Applied Sciences Vol. 356,
Kluwer Academic Publishers, Dordrecht 1999. The anisotropy ratio
$\Gamma=5.3$ measured at $T_{\rm c}$ means that a field of 16 T,
$B\|ab$, produces the same quantitative change on the specific
heat anomaly at $T_{\rm c}$ as a field of 3 T, $B\|c$.
\bibitem{41} F. Gygi and M. Schl\"uter, Phys. Rev. B 43, 7609 (1991).
\bibitem{42} D. Sanchez, A. Junod, J.Muller, H. Berger and F. L\'evy,
 Physica C 204, 167 (1995).
\bibitem{43} E. Schachinger, private communication.
\bibitem{44} W.N. Hardy, S. Kamal and D.A. Bonn, in: "The Gap Symmetry and Fluctuations
in High-\tc~Superconductors", eds. J. Bok, G. Deutscher, D. Pavuna
and S.A. Wolf, Plenum Press, New York 1998, p. 373.
\bibitem{45} A.P. Ramirez, Phys. Lett. A 211, 59 (1996).
\bibitem{46} M. Hedo, Y. Inada, E. Yamamoto, Y. Haga, Y. Onuki, Y. Aoki,
 T.D. Matsuda, H. Sato and S. Takahashi, J. Phys. Soc. Jpn. 67, 272 (1998).
\bibitem{47} J.E. Sonier, M.F. Hunderly, J.D. Thompson and J.W. Brill,
 Phys. Rev. Lett. 82, 4914 (1999).
\bibitem{48} M. Ichioka, A. Hasegawa and K. Machida, Phys. Rev. B 59, 184 (1999).
\bibitem{49} I. Maggio-Aprile, private communication.
\bibitem{50} B. Revaz, J.-Y. Genoud, A. Junod, K. Neumaier, A. Erb and E.
Walker, Phys. Rev. Lett. 80, 3364 (1998).
\bibitem{51} H. Won and K. Maki, cond-mat/0004105.
\bibitem{52} M. Franz and Z. Tesanovic, Phys. Rev. Lett.
80, 4763 (1998).
\bibitem{53} M. Franz and Z. Tesanovic, Phys. Rev. Lett. 84,
554 (2000).
\bibitem{54} O. Vafek, A. Melikyan, M. Franz and Z. Tesanovic,
cond-mat/0007296.
\bibitem{55} Luca Marinelli, B. I. Halperin and S. H. Simon,
cond-mat/0001406.
\end{references}
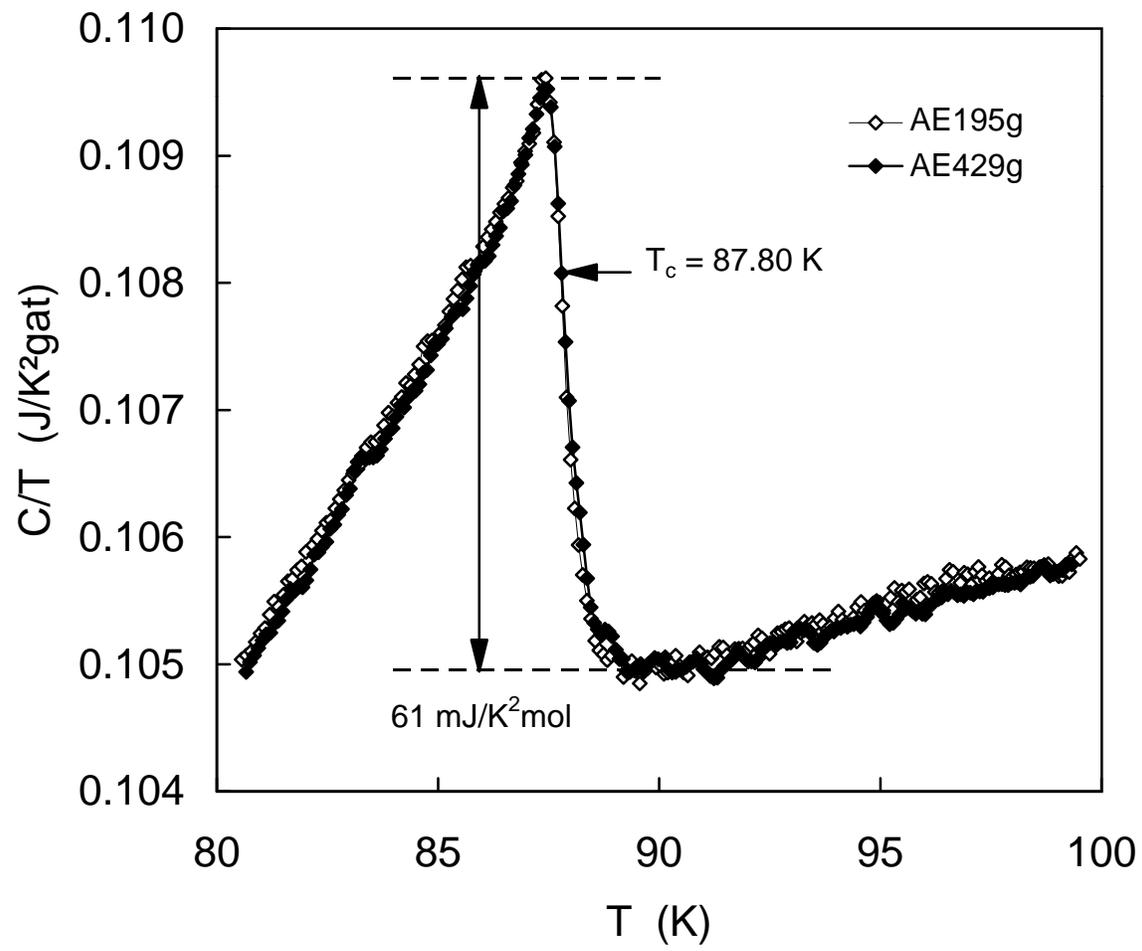
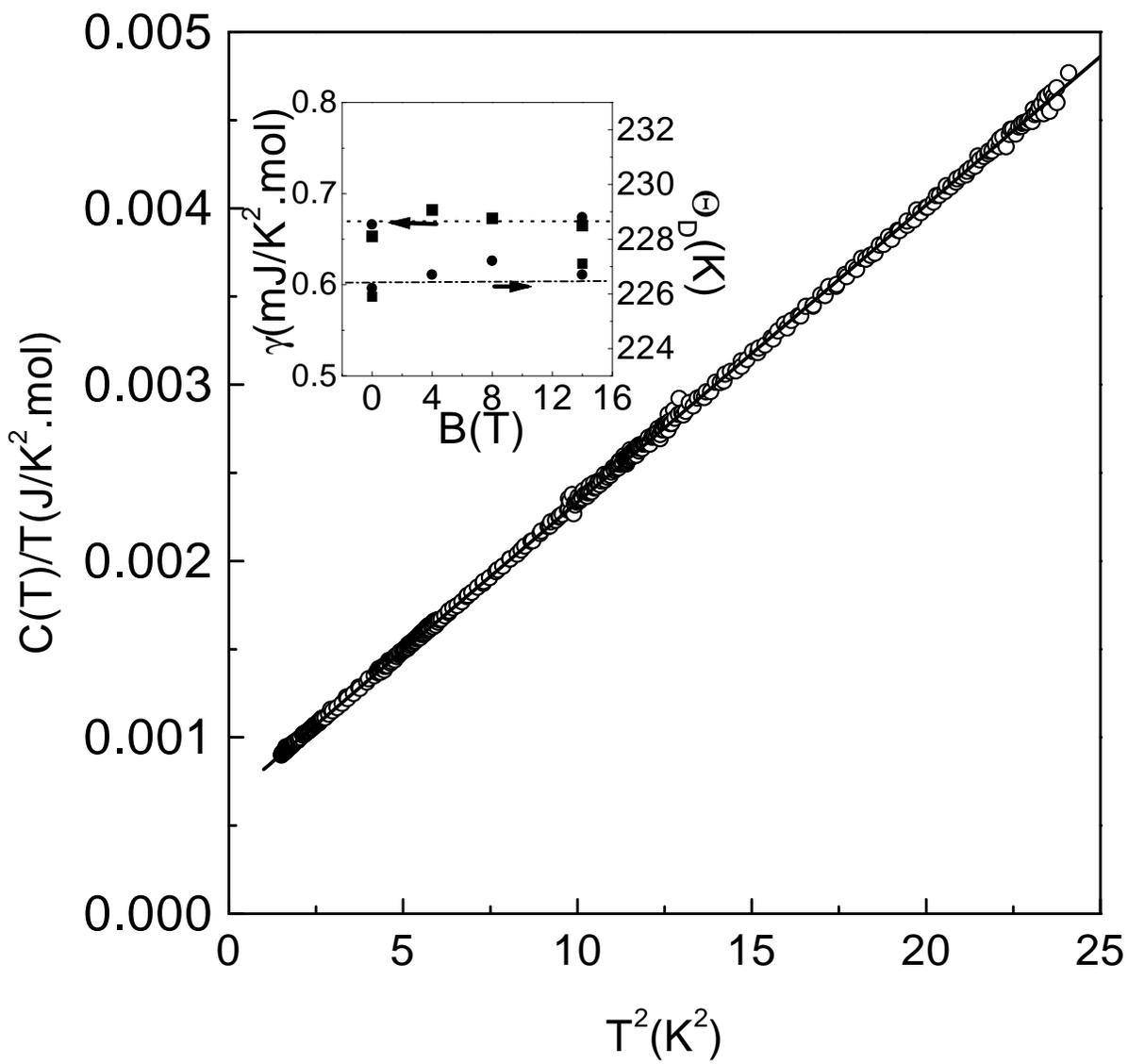
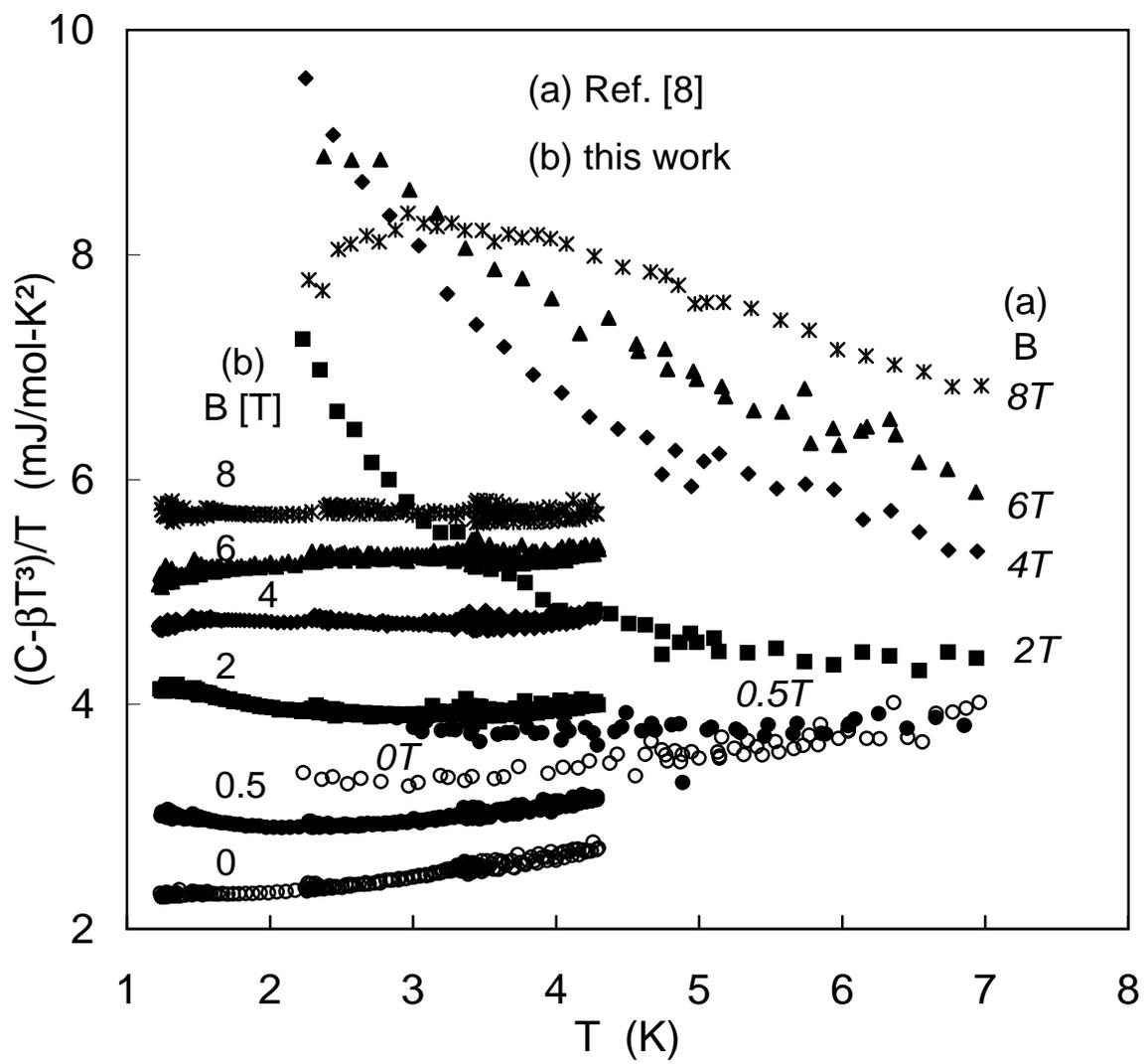
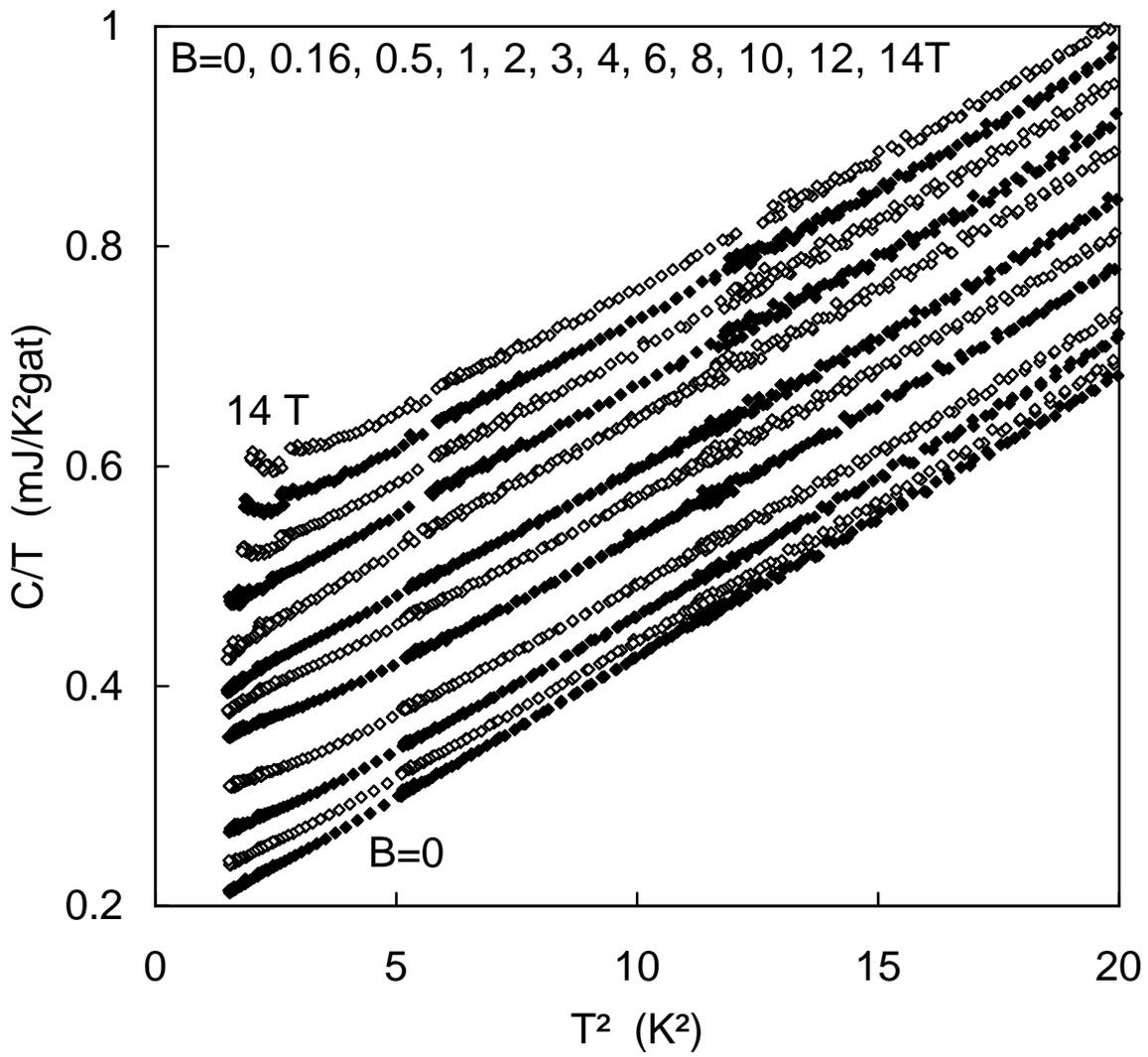
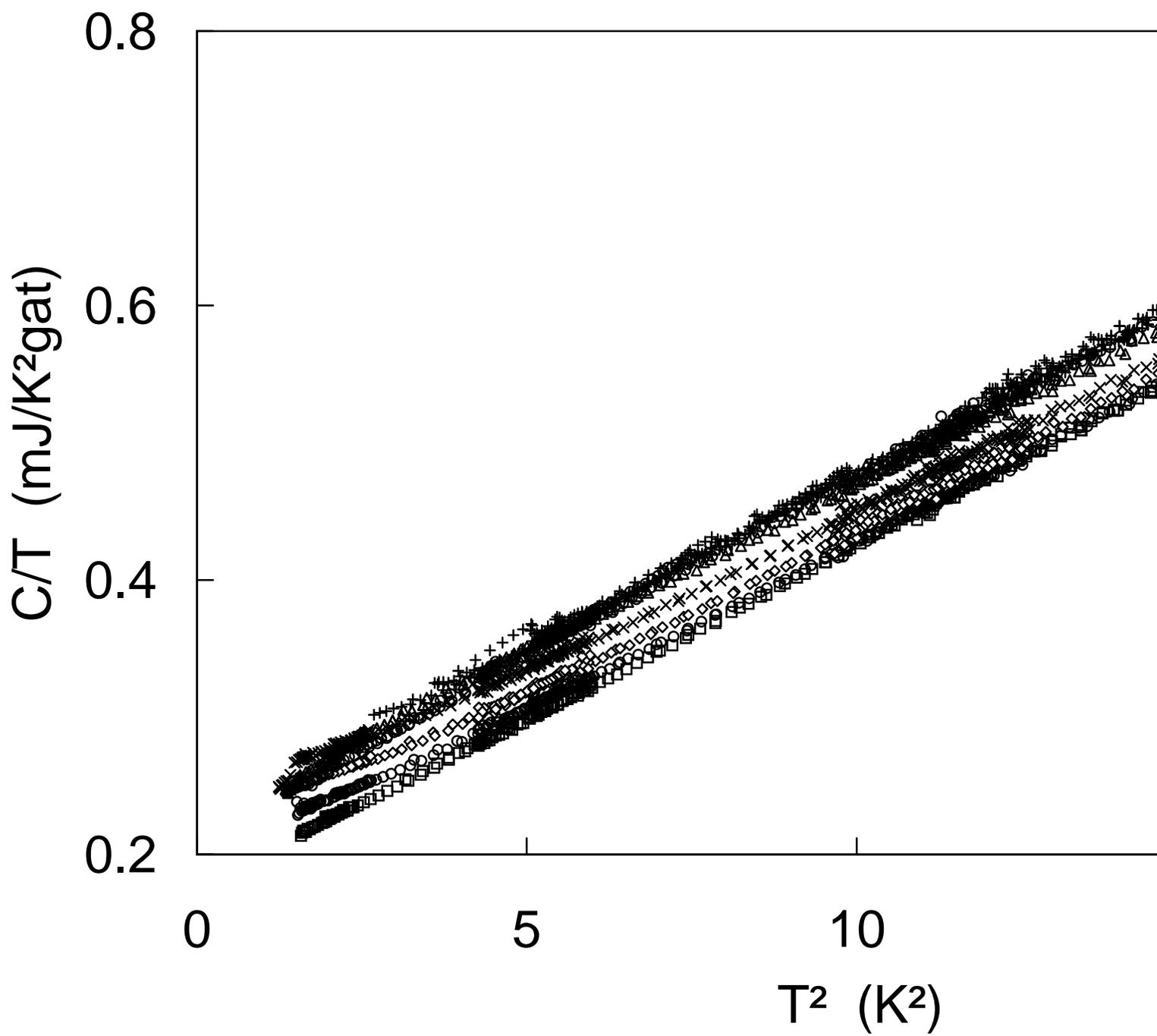
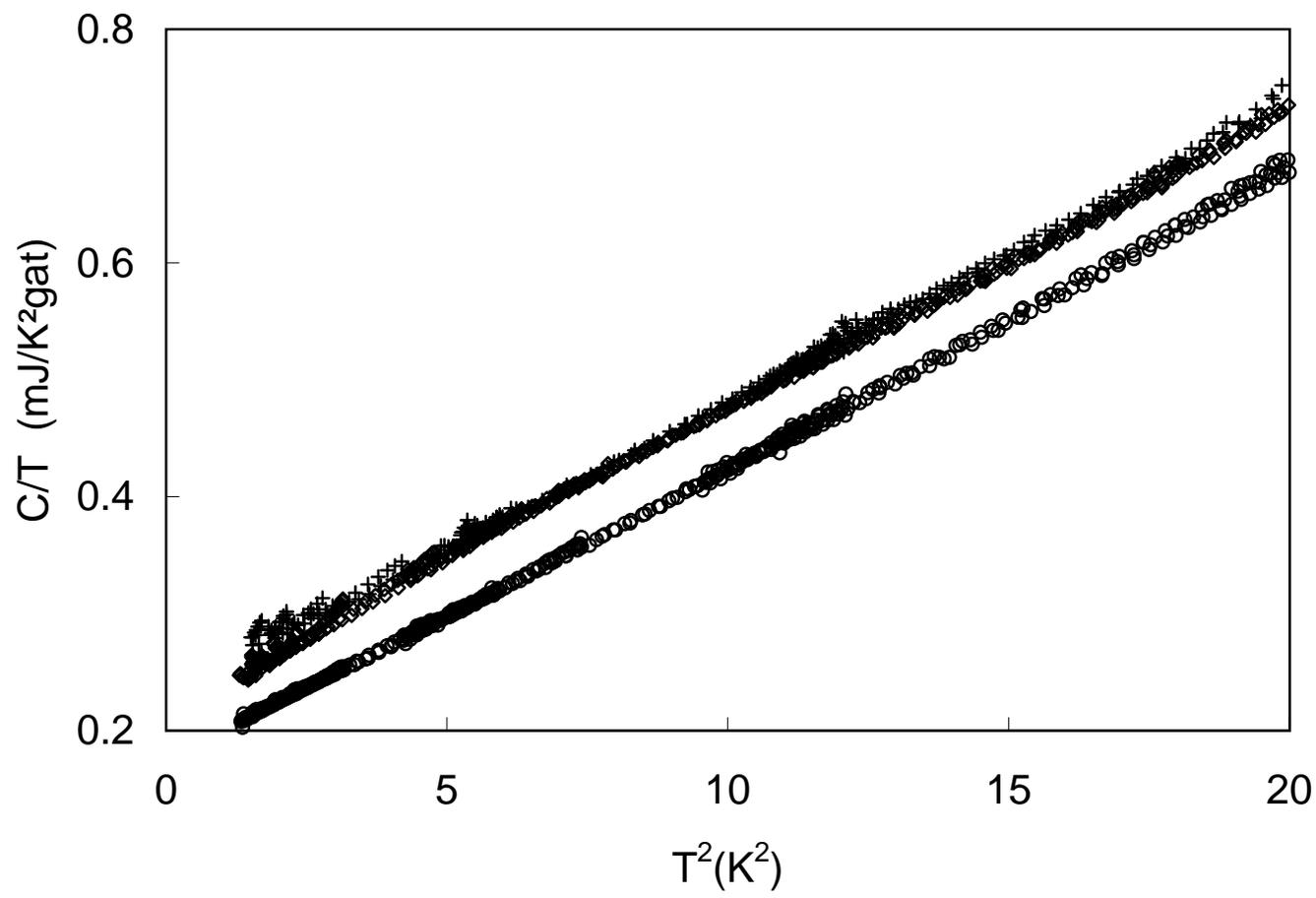
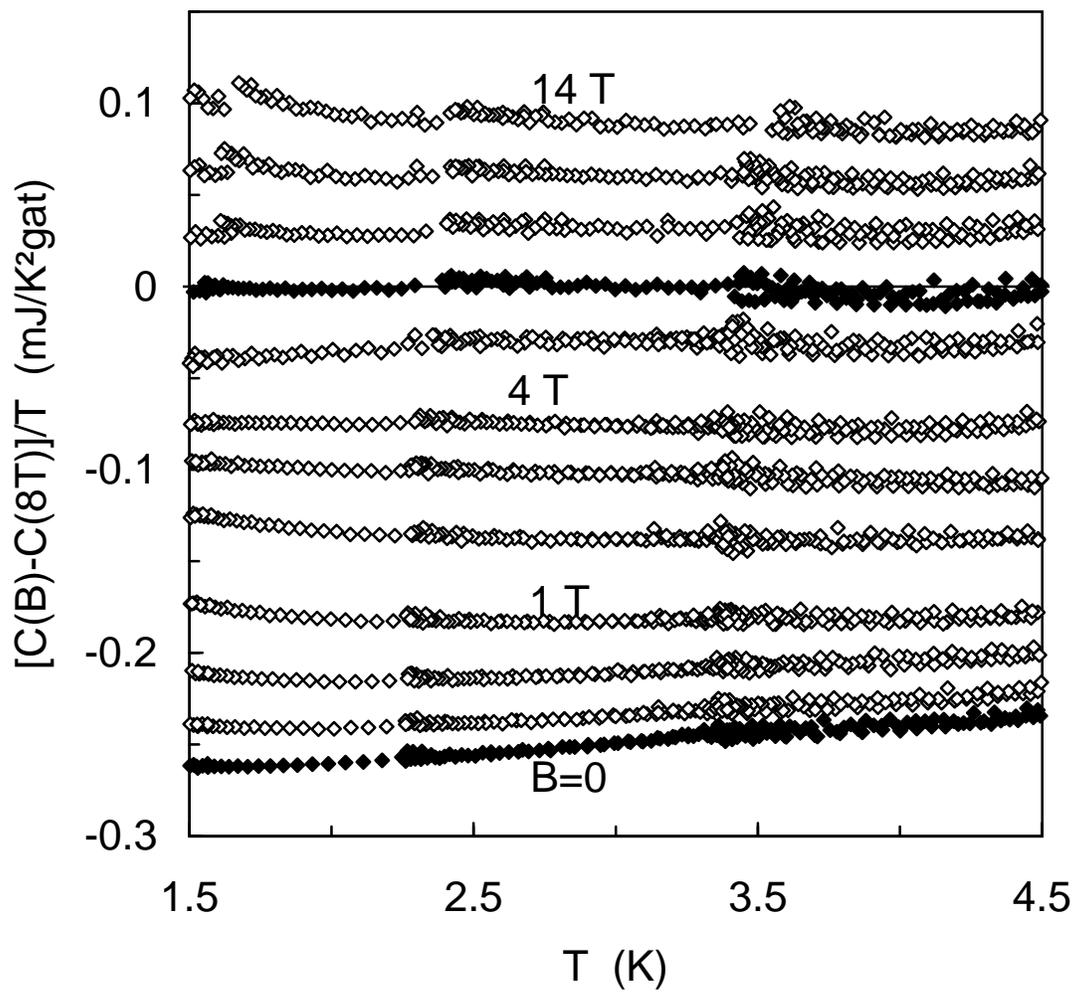
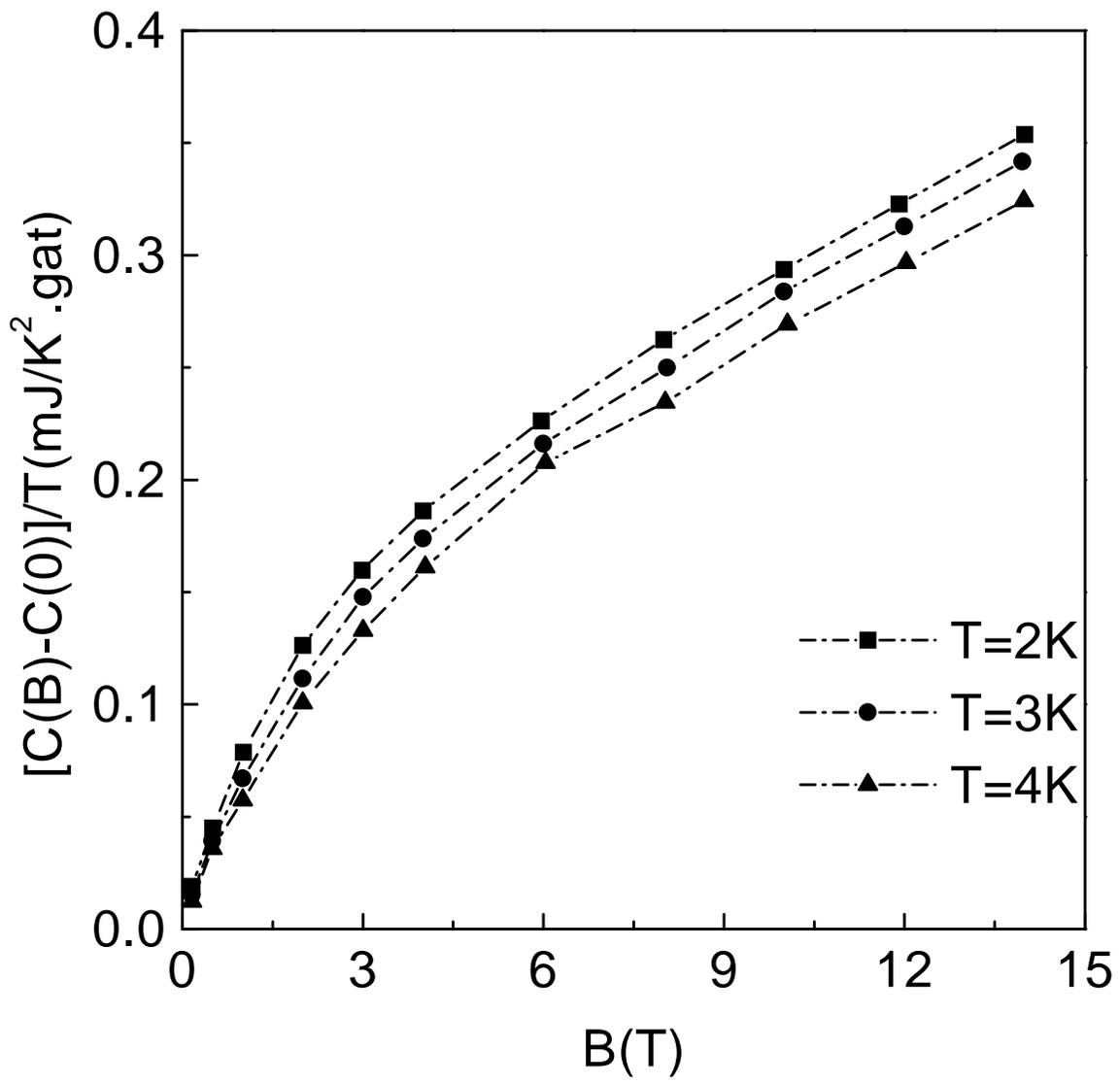
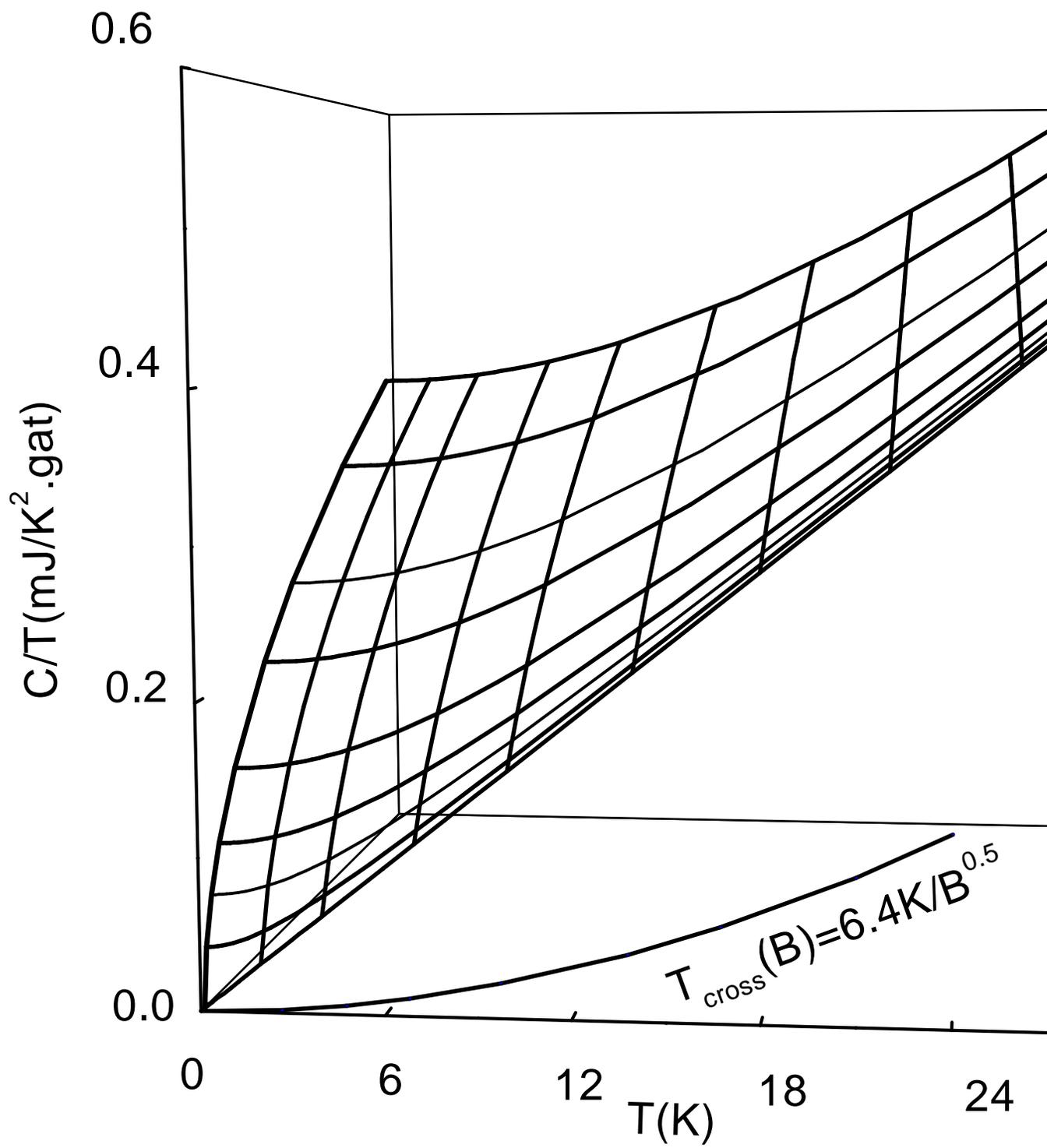
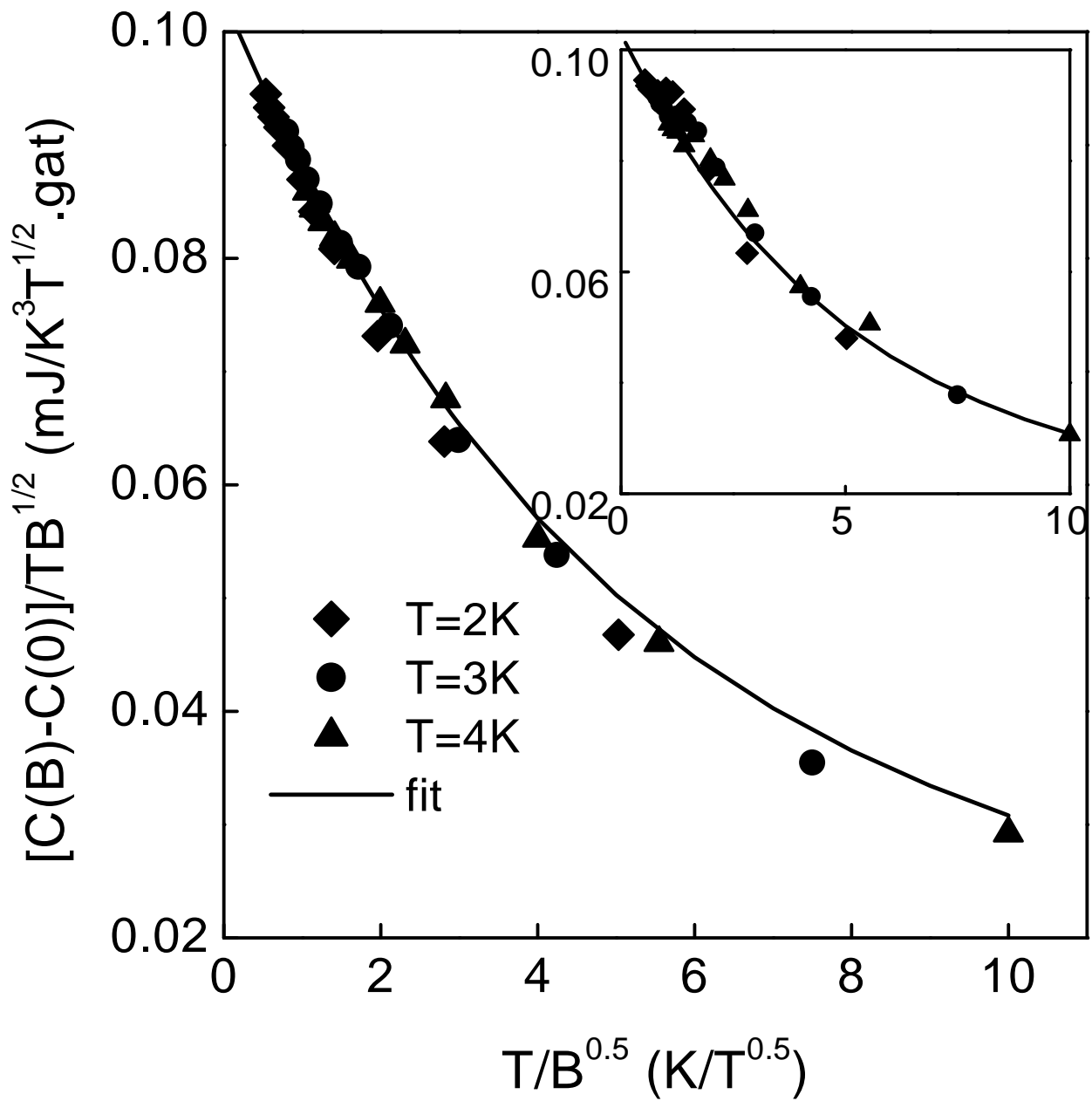
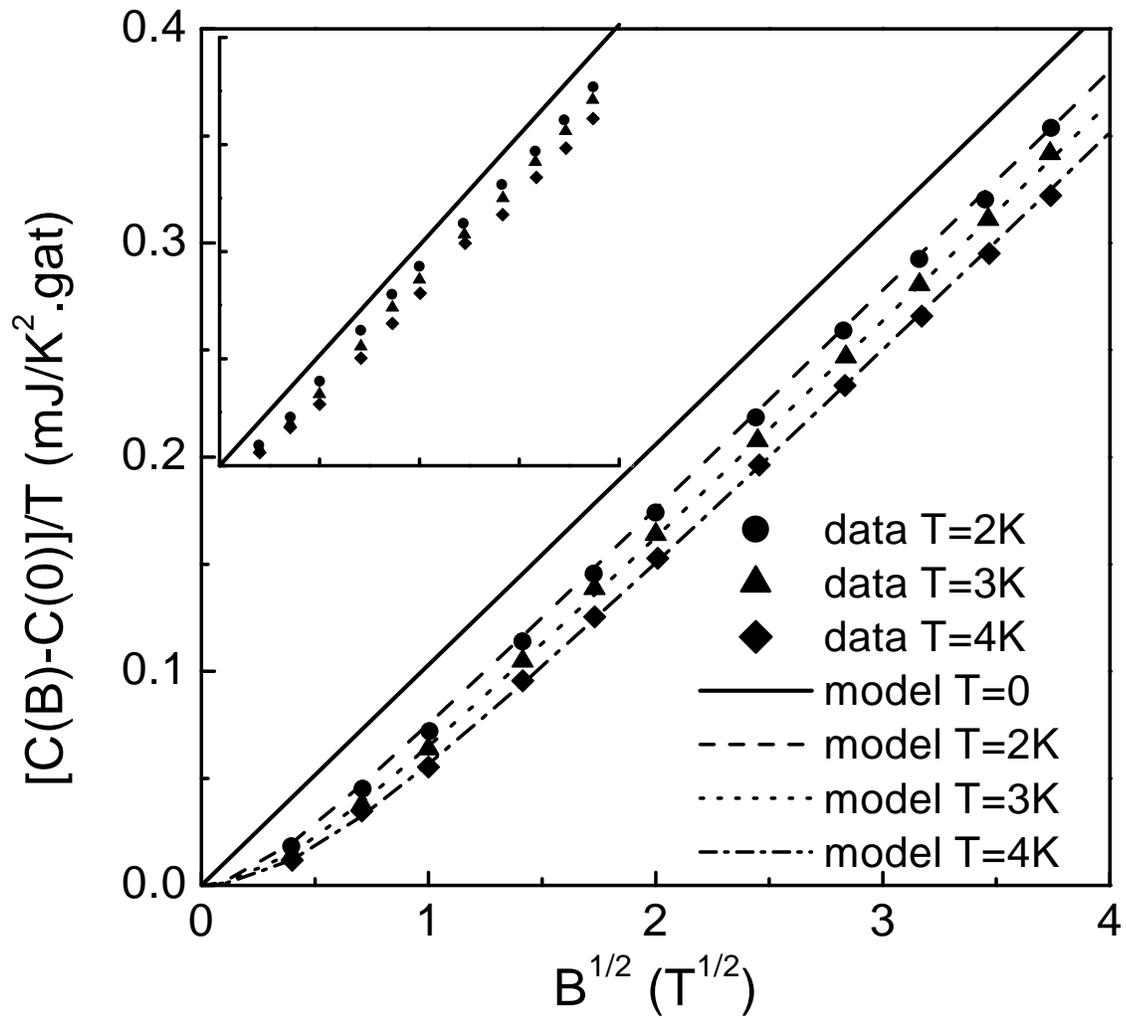
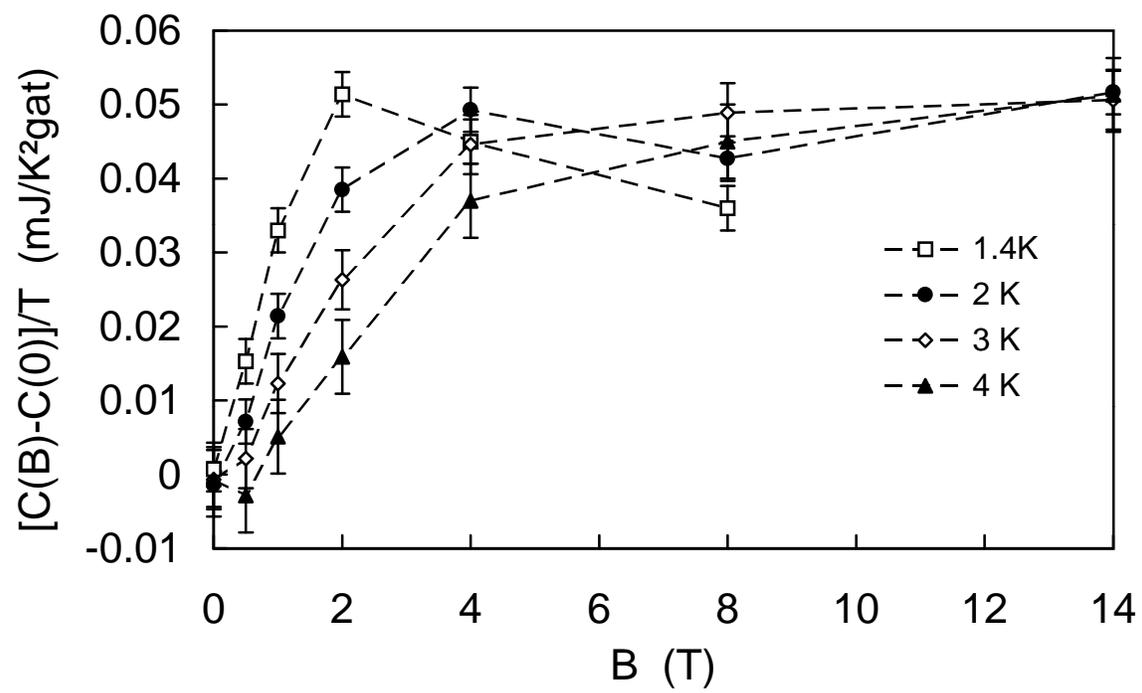
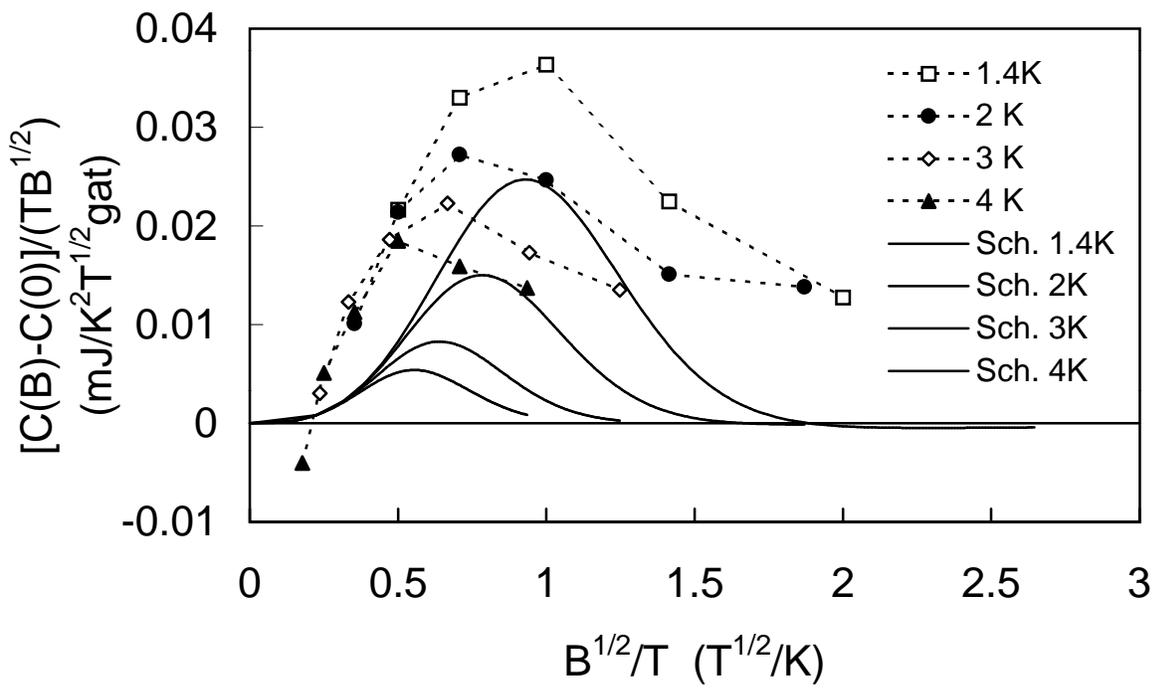
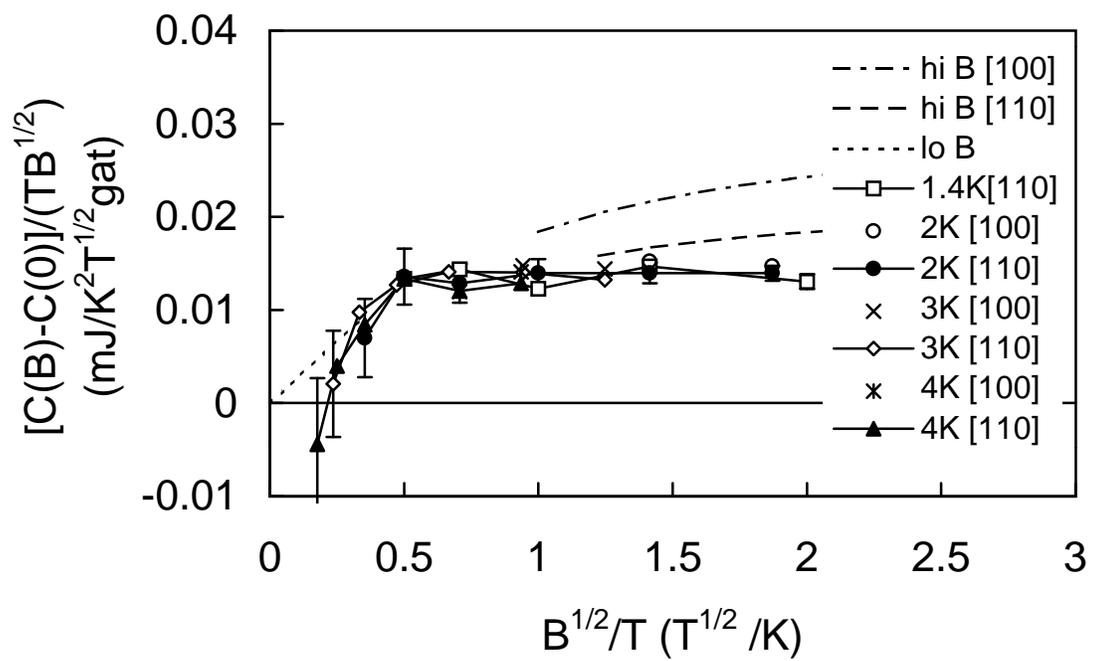
\end{document}